\documentclass[journal]{IEEEtran}

\usepackage{graphicx}
\usepackage{xcolor}
\usepackage{amssymb}
\usepackage{multirow}
\usepackage{multicol}
\usepackage{diagbox}
\usepackage{tabularx}
\usepackage[cmex10]{amsmath}
\usepackage{cite}
\usepackage{url}
\usepackage[normalem]{ulem}
\usepackage{pifont}
\usepackage{dsfont}
\usepackage[absolute,showboxes]{textpos}

\TPMargin{3pt}

\newcommand{\copyrightstatement}{
    \begin{textblock}{0.84}(0.08,0.953) % tweak here: {box width}(left position, bottom position)
         \noindent
         \footnotesize
         \copyright 2023 IEEE. Personal use of this material is permitted. Permission from IEEE must be obtained for all other uses, in any current or future media, including reprinting/republishing this material for advertising or promotional purposes, creating new collective works, for resale or redistribution to servers or lists, or reuse of any copyrighted component of this work in other works. DOI: 10.1109/TBC.2023.3275363
    \end{textblock}
}
\begin{document}
\copyrightstatement
% paper title
% Titles are generally capitalized except for words such as a, an, and, as,
% at, but, by, for, in, nor, of, on, or, the, to and up, which are usually
% not capitalized unless they are the first or last word of the title.
% Linebreaks \\ can be used within to get better formatting as desired.
% Do not put math or special symbols in the title.
\title{Precoding Based Downlink OAM-MIMO Communications with Rate Splitting}
%
% author names and IEEE memberships
% note positions of commas and nonbreaking spaces ( ~ ) LaTeX will not break
% a structure at a ~ so this keeps an author's name from being broken across
% two lines.
% use \thanks{} to gain access to the first footnote area
% a separate \thanks must be used for each paragraph as LaTeX2e's \thanks
% was not built to handle multiple paragraphs
%
\author{Ruirui~Chen,
        Jinyang~Lin,
        Beibei~Zhang,
        Yu~Ding,
        Keyue~Xu
        %and~Jane~Doe,~\IEEEmembership{Life~Fellow,~IEEE}% <-this % stops a space
\thanks{This work is supported in part by Natural Science Foundation of Jiangsu Province under Grant BK20200650, in part by National Natural Science Foundation of China under Grant 62071472, in part by China Postdoctoral Science Foundation under Grant 2019M660133, in part by Program for ``Industrial IoT and Emergency Collaboration'' Innovative Research Team in CUMT under Grant 2020ZY002, in part by Fundamental Research Funds for the Central Universities under Grant 2019QNB01 and 2020ZDPY0304. (Corresponding author: Ruirui~Chen.)
\par Ruirui Chen, Jinyang Lin, Beibei~Zhang, Yu~Ding and Keyue Xu are with School of Information and Control Engineering, China University of Mining and Technology, Xuzhou, 221116, China (emails: rrchen@cumt.edu.cn, linjinyang@cumt.edu.cn, lb15060029@cumt.edu.cn, yding@cumt.edu.cn, kyxu@cumt.edu.cn). Beibei~Zhang is also with Jiangsu Automation Research Institute, Lianyungang, 222061, China.}
%\par Tao Li is with the State Key Laboratory of Integrated Services Networks, Xidian University, Xi¡¯an 710071, China (e-mail: taoli@xidian.edu.cn).}

% <-this % stops a space
%\thanks{J. Doe and J. Doe are with Anonymous University.}% <-this % stops a space
%\thanks{Manuscript received April 19, 2005; revised August 26, 2015.}
}
% note the % following the last \IEEEmembership and also \thanks -
% these prevent an unwanted space from occurring between the last author name
% and the end of the author line. i.e., if you had this:
%
% \author{....lastname \thanks{...} \thanks{...} }
%                     ^------------^------------^----Do not want these spaces!
%
% a space would be appended to the last name and could cause every name on that
% line to be shifted left slightly. This is one of those "LaTeX things". For
% instance, "\textbf{A} \textbf{B}" will typeset as "A B" not "AB". To get
% "AB" then you have to do: "\textbf{A}\textbf{B}"
% \thanks is no different in this regard, so shield the last } of each \thanks
% that ends a line with a % and do not let a space in before the next \thanks.
% Spaces after \IEEEmembership other than the last one are OK (and needed) as
% you are supposed to have spaces between the names. For what it is worth,
% this is a minor point as most people would not even notice if the said evil
% space somehow managed to creep in.

% The paper headers
\markboth{IEEE Transactions on Broadcasting, VOL. 69, NO. 4, DECEMBER 2023}%
{Shell \MakeLowercase{\textit{et al.}}: Bare Demo of IEEEtran.cls for IEEE Journals}
% The only time the second header will appear is for the odd numbered pages
% after the title page when using the twoside option.
%
% *** Note that you probably will NOT want to include the author's ***
% *** name in the headers of peer review papers.                   ***
% You can use \ifCLASSOPTIONpeerreview for conditional compilation here if
% you desire.

% If you want to put a publisher's ID mark on the page you can do it like
% this:
%\IEEEpubid{0000--0000/00\$00.00~\copyright~2015 IEEE}
% Remember, if you use this you must call \IEEEpubidadjcol in the second
% column for its text to clear the IEEEpubid mark.

% use for special paper notices
%\IEEEspecialpapernotice{(Invited Paper)}

% make the title area
\maketitle

% As a general rule, do not put math, special symbols or citations
% in the abstract or keywords.
\begin{abstract}
Orbital angular momentum (OAM) and rate splitting (RS) are the potential key techniques for the future wireless communications. As a new orthogonal resource, OAM can achieve the multifold increase of spectrum efficiency to relieve the scarcity of the spectrum resource, but how to enhance the privacy performance imposes crucial challenge for OAM communications.
RS technique divides the information into private and common parts, which can guarantee the privacies for all users.
In this paper, we integrate the RS technique into downlink OAM-MIMO communications, and study the precoding optimization to maximize the sum capacity. First, the concentric uniform circular arrays (UCAs) are utilized to construct the downlink transmission framework of OAM-MIMO communications with RS. Particularly, users in the same user pair utilize RS technique to obtain the information and different user pairs use different OAM modes. Then, we derive the OAM-MIMO channel model, and formulate the sum capacity maximization problem. Finally, based on the fractional programming, the optimal precoding matrix is obtained to maximize the sum capacity by using quadratic transformation.
Extensive simulation results show that by using the proposed precoding optimization algorithm, OAM-MIMO communications with RS can achieve higher sum capacity than the traditional communication schemes.
\end{abstract}

% Note that keywords are not normally used for peerreview papers.

% For peer review papers, you can put extra information on the cover
% page as needed:
% \ifCLASSOPTIONpeerreview
% \begin{center} \bfseries EDICS Category: 3-BBND \end{center}
% \fi
%
% For peerreview papers, this IEEEtran command inserts a page break and
% creates the second title. It will be ignored for other modes.
\IEEEpeerreviewmaketitle

\section{Introduction}
\subsection{Background}

\IEEEPARstart{T}{he} rapid growing demand of spectrum efficiency in the big data era brings great challenges to existing traditional communication technologies \cite{1}.
%drives wireless communication technology development\cite{1,2}. To meet the requirement, more and more high frequency bands such as millimeter wave and terahertz bands are being licensed [2].
Due to the scarcity of the spectrum resource, it becomes increasingly important to search for the new orthogonal resource \cite{2a,2}.
Therefore, orbital angular momentum (OAM), which can be considered as a new orthogonal resource, provides a new solution to the multifold increase of spectrum efficiency for the future wireless communications \cite{3a,3}.
The vortex waves with different OAM modes are orthogonal to each other, and thus the infinite OAM modes can be used to significantly improve the spectrum efficiency in theory \cite{4,5}.
However, vortex waves with non-integer OAM modes are difficult to decode, and vortex waves with high OAM modes result in serious attenuation and beam divergence, which limits the user access number of the OAM communications \cite{6}.
Therefore, improving the access capability is necessary for the application of OAM communications into the multi-user scenario.

To serve large number of users, OAM communications can utilize the space division multiple access (SDMA) technique or the non-orthogonal multiple access (NOMA) technique \cite{7,8}.
By using SDMA technique, each user directly decodes the information by treating undesired signals as interferences, which leads to the rate saturation even with excess transmit power when the transmit antenna number is insufficient or a resource block is used by multiple users \cite{9}.
It is difficult for SDMA technique to support the future high-load traffic scenarios, where there are high-rate services caused by many users at the same time \cite{10}.
NOMA technique performs successive interference cancellation (SIC) at the receiver, which can realize the information transmission at the same time and frequency \cite{11}.
In NOMA transmission, only the last user can guarantee the private information, which is not decoded by other users \cite{12}.
%However, both NOMA and SDMA can only securely transmit private message for a specific user rather than all users \cite{13}.

Rate splitting (RS) technique combines the advantages of SDMA and NOMA by the flexible balance between the interference and desired signal \cite{13,14}.
By dividing the transmission information into private message and common message , RS technique can serve multiple users at the same time and frequency \cite{15}. All users use SIC technique to remove the common message and decode their own private message \cite{16}.
Thus, RS technique decodes part of the interference and treats part of the interference as noise \cite{17}, which avoids completely treating interference as noise and decoding interference \cite{18}.
In addition, RS technique can guarantee the private transmission of all users and be applied to high-load traffic scenarios for serving large number of users \cite{19}.

In this paper, we optimize the precoding matrix to maximize the sum capacity of the downlink OAM-MIMO communications with RS, which can guarantee the privacy performance for all users.

\subsection{Related Works}
In this paper, RS technique is introduced to the downlink OAM-MIMO communications, and precoding matrix is optimized to maximize the sum capacity.
We will divide previous contributions related with this work into two aspects, i.e., precoding design for wireless communications with RS and beam steering design for OAM communications.

\subsubsection{Precoding Design for Wireless Communications with RS}
As early as 1981, T. Han and K. Kobayashir proposed the RS technique \cite{20}.
Based on linear precoding and SIC at the receiver, RS technique decodes part of the interference and treats the remaining part of the interference as noise \cite{21}.
It outperforms SDMA and NOMA techniques in terms of network load and user privacy \cite{22}.
Dividing message into a private part and a common part is the key of the precoding design for wireless communications with RS \cite{23}.
In \cite{24}, RS technique was proposed to be utilized in the multigroup multicast precoding design.
The authors of \cite{25} investigated RS technique to improve the energy efficiency of the multi-user multicast MISO system.
%Compared with NOMA and SDMA, RSMA is demonstrated to further exploit spatial dimensions.
Furthermore, RS technique achieves higher energy efficiency and requires less channel state information feedback \cite{26,27,28,29}.
The work in \cite{30} indicates that RS technique is more robust to the impact factors such as channel disparity, channel orthogonality, network load and channel quality.
These results demonstrate that the RS technique obtains better performance compared with other access techniques.

\subsubsection{Beam Steering Design for OAM Communications}
%The application of OAM to wireless communications is expected to break the capacity theoretical limitation of existing communication systems \cite{31,32}.
%However, there are still some technical challenges for the practical application of OAM communications.
Although vortex wave carrying OAM is a promising beyond 5G technique \cite{31,32}, there are still some technical challenges for the practical application of OAM communications.
One challenge is that OAM communications require perfect alignment between the transmit and receive antenna arrays \cite{33}. For OAM narrow-band communication system, the authors in \cite{34} drawn a conclusion that if the antennas are not aligned, the system performance quickly deteriorates. To solve this problem, the authors of \cite{35} studied the effect of misalignment on the performance of OAM broadband system and proposed a beam steering scheme for the misaligned multi-mode OAM communications. In \cite{36}, the authors proposed a hybrid position phase beam steering (HPPBS) scheme to overcome the performance deterioration caused by the misalignment of uniform circular array (UCA) based OAM communications. The authors of \cite{37} investigated the effect of the misalignment on channel capacity, and the proposed beam steering method can avoid the performance deterioration caused by misalignment. The proposed beam steering method in \cite{38} can eliminate inter-mode interferences induced by the misalignment error between the transmit and receive UCAs.
In misaligned and incomplete receive UCA scenario, a partial receive scheme based on beam steering was proposed for the non-ideal OAM-MIMO communications to address the performance degradation caused by OAM beam divergence and misalignment \cite{39}. Considering the size and cost of UCA, the authors in \cite{40} developed an efficient technique for beam steering of arbitrary sizes of circular arrays based on OAM modes. For OAM communications, the authors of \cite{41} presented a novel practical realization of field eigenmode beamforming in the azimuthal domain. To avoid multiple radio frequency chains and costly analog devices, a beam steering method was proposed in \cite{42} to generate arbitrary-order OAM mode.
The authors of \cite{43} utilized the spherical conformal array antenna to achieve beam steering of vortex waves, which can be applied in OAM communications.
In \cite{44}, the authors proposed beam steering method to solve the aperture alignment of the transmitter and receiver in OAM communications. Through superposition of vortex waves with different OAM modes, the authors of \cite{45} achieved single and multiple beam steerings in the transverse direction.

\subsection{Motivations}

To enhance the privacy performance of OAM communications, we are motivated to study the downlink OAM-MIMO communications with RS, where multiple vortex wave signals are simultaneously transmitted and privacy transmission can be guaranteed for all users. Then, the challenges are discussed as follows.
\begin{itemize}
  \item \emph{Limited OAM Mode:} Not all OAM modes can be used to achieve the high-capacity OAM communications. For the non-integer OAM mode, it is difficult to decode the vortex wave signal due to the nonuniform phase distribution. Furthermore, OAM communications with the high-order OAM mode suffers from severe divergence and attenuation. In practice, OAM communications mainly utilize the low-order integer OAM mode, thus dissatisfying the large number of user access requirements, which seriously limits its application in the multi-user scenario.
  \item \emph{Lack of Precoding Algorithm:} The vortex wave signal is transmitted by the UCA, each element of which generates the phase shift to produce different OAM modes. Then, the transmission information is divided into private and common parts. The integration of RS technique into OAM communications complicates the transmission framework, which results in the difficult formulation of sum capacity optimization problem. Furthermore, it is difficult to obtain the optimal precoding matrix for non-convex sum capacity maximization problem.
\end{itemize}

\subsection{Main Contributions}

To address the challenges, the sum capacity optimization problem is formulated for the downlink OAM-MIMO communications with RS. However, the formulated optimization problem is non-convex. We utilize the fractional programming (FP) with quadratic transformation to obtain the optimal precoding matrix.

The main contributions of this paper are summarized as follows.
\begin{itemize}
  \item By integrating RS into OAM communications, different user pairs use different OAM modes and users in the same user pair utilize RS technique to obtain the information. The transmission framework is presented for the downlink OAM-MIMO communications with RS, where the UCA is used to transmit the vortex wave signal.
  \item The channel model is derived based on the UCA. The non-convex optimization problem is formulated to maximize the sum capacity of the downlink OAM-MIMO communications with RS. Based on the FP, the optimal precoding matrix is obtained by using the quadratic transformation.
  \item Simulation results demonstrate the capacity of users in the same user pair and analyze the impact of the distance on the downlink OAM-MIMO communications with RS. Furthermore, RS technique is superior to the other traditional access schemes.
\end{itemize}

 The rest of this paper is organized as follows. Section II gives the system model. In Section III, the transmission framework is proposed for the downlink OAM-MIMO communications with RS. Section IV derives the channel model and formulates the sum capacity maximization problem. In Section V, we derive the optimal precoding matrix based on the FP. Section VI gives the simulation results to evaluate the proposed precoding optimization algorithm for the downlink OAM-MIMO communications with RS. Finally, the paper concludes with Section VII.

\section{System Model}

This paper considers the downlink OAM-MIMO communications with RS, which consist of multiple user devices (UDs) and a base station (BS). The UDs are denoted by ${\rm{U}}{{\rm{D}}_i}$, where $i = \{ 1,{\kern 1pt} {\kern 1pt} {\kern 1pt} 2, \ldots ,2k - 1,2k, \ldots ,2K - 1,2K\}$. Based on RS technique, the transmission information is partitioned into private message and common message. The UDs should be divided into pairs, which can achieve better privacy performance. In this paper, we choose two UDs as one pair. The reason is that the two user pair is the most general case for the reader to understand our work and easy to be realized in the practical wireless communications. Note that although two user pair is chosen, our work can be extended to the user pair with more users. Furthermore, we choose the two users, which have the nearest distance, as one user pair.

Thus, the $k$-th user pair $({\rm{U}}{{\rm{D}}_{2k-1}}, {\rm{U}}{{\rm{D}}_{2k}})$ can complete the wireless communications by dividing their information into private message and common message. The information of ${\rm{U}}{{\rm{D}}_{2k-1}}$ and ${\rm{U}}{{\rm{D}}_{2k}}$ can be respectively partitioned into $\left\{ {F_{2k-1}^{\rm{p}},F_{2k-1}^{\rm{c}}} \right\}$ and $\left\{ {F_{2k}^{\rm{p}},F_{2k}^{\rm{c}}} \right\}$. Furthermore, the private messages $F_{2k-1}^{\rm{p}}$ and $F_{2k}^{\rm{p}}$ are encoded separately into transmit signals ${\hat x_{2k-1}}$ and ${\hat x_{2k}}$, which are respectively decoded by ${\rm{U}}{{\rm{D}}_{2k-1}}$ and ${\rm{U}}{{\rm{D}}_{2k}}$. The common messages $F_{2k-1}^{\rm{c}}$ and $F_{2k}^{\rm{c}}$ are encoded into transmit signal ${\hat x_{2k - 1,2k}}$, which are decoded by both ${\rm{U}}{{\rm{D}}_{2k-1}}$ and ${\rm{U}}{{\rm{D}}_{2k}}$. The $k$-th user pair $({\rm{U}}{{\rm{D}}_{2k-1}}, {\rm{U}}{{\rm{D}}_{2k}})$ can use the carrier with OAM mode ${l_k}$. In other words, each user pair uses one OAM mode. From the theoretical perspective, there exist infinite user pairs due to infinite OAM modes.

OAM communications utilize the UCA to transmit/receive vortex wave signal in general. The UCA based system model is demonstrated in Fig. 1. Each UD is equipped with one UCA to receive the vortex wave signal. BS has the transmit concentric UCAs, which include different antenna elements on different UCAs with the same center. In the downlink OAM-MIMO communications with RS, BS assigns the $k$-th transmit UCA to serve the $k$-th user pair $({\rm{U}}{{\rm{D}}_{2k-1}}, {\rm{U}}{{\rm{D}}_{2k}})$. The $k$-th transmit UCA consists of $M_k$ antenna elements, which are denoted by $m_k \in \textbf{\rm{M}}=\{1,2, \ldots ,M_k\}$. The center of the $k$-th transmit UCA is at the location ${\rm{(0, 0, 0)}}$ in the 3D cylindrical coordinate system. The $z$-axis is the line across the center of transmit concentric UCAs and perpendicular to the transmit concentric UCAs. The receive UCA of ${\rm{U}}{{\rm{D}}_i}$ has ${N_i}$ antenna elements, which are represented by ${n_i} \in \textbf{\rm{N}} = \{1,2, \ldots ,{N_i}\}$. The distance from the ${i}$-th UCA center of ${\rm{U}}{{\rm{D}}_i}$ to the $k$-th transmit UCA center of BS is denoted by ${d_{k,i}}$.
\begin{figure}[thp]
\centering
\includegraphics[height=2.8in,width=8.8cm]{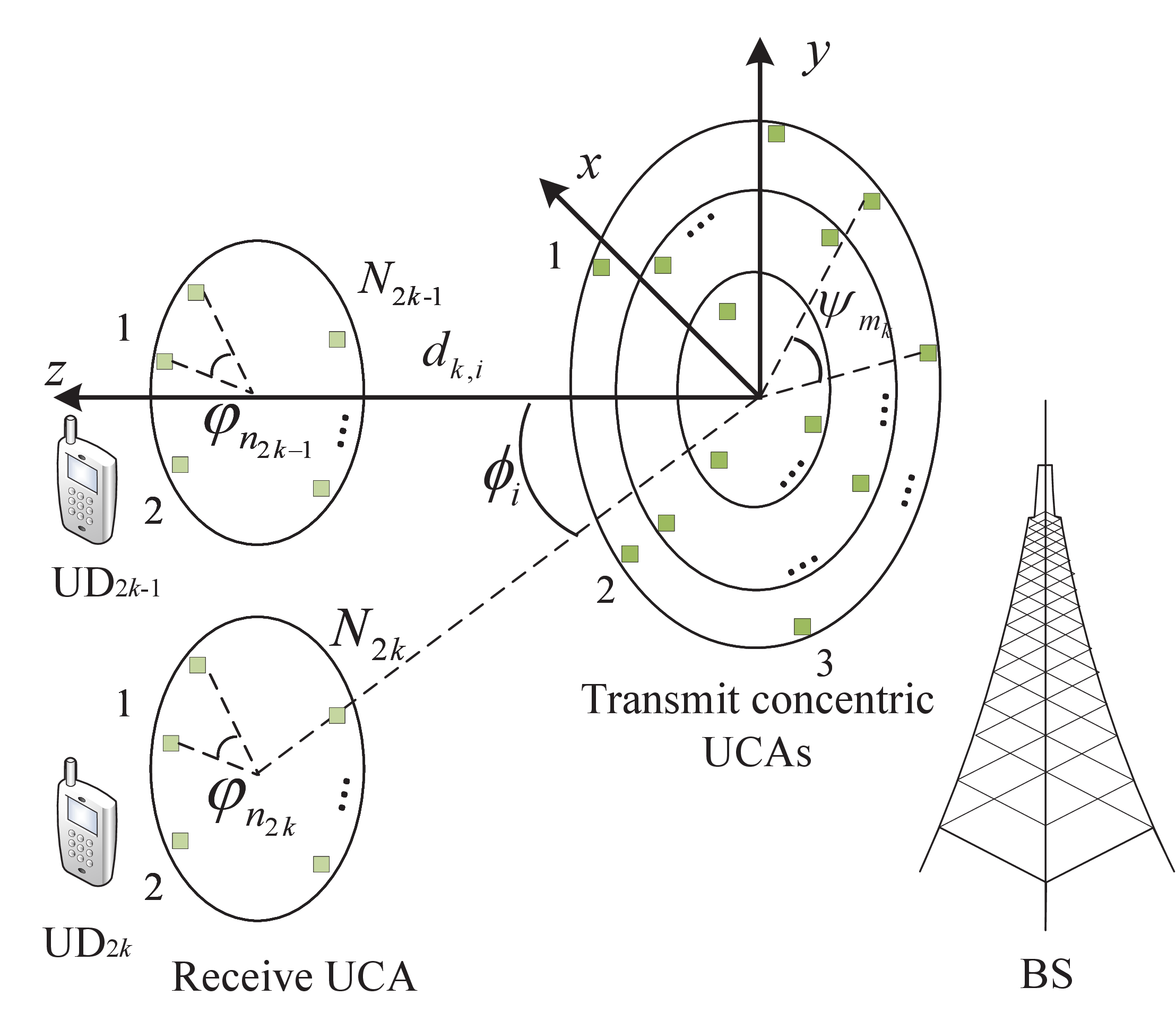}
\caption{Downlink OAM-MIMO communications with RS}
\label{fig:side:c}
\intextsep=1pt plus 3pt minus 1pt
\end{figure}

\section{Transmission Framework of Downlink OAM-MIMO Communications with RS}
In the downlink OAM-MIMO communications with RS, the $k$-th UCA transmits the RS signal ${x_k} = p_{2k - 1}^{\rm{p}}\hat x_{2k - 1}^{\rm{p}} + p_{2k}^{\rm{p}}\hat x_{2k}^{\rm{p}} + p_{2k - 1,2k}^{\rm{c}}\hat x_k^{\rm{c}}$ to the $k$-th user pair $({\rm{U}}{{\rm{D}}_{2k - 1}}, {\rm{U}}{{\rm{D}}_{2k}})$ by using the carrier with OAM mode ${l_k}$, where $p_{2k - 1,2k}^{\rm{c}}$ is the precoding vector of common message, $p_{2k - 1}^{\rm{p}}$ and $p_{2k}^{\rm{p}}$ are respectively the precoding vectors of private messages for ${\rm{U}}{{\rm{D}}_{2k - 1}}$ and ${\rm{U}}{{\rm{D}}_{2k}}$.
Moreover, the total transmit power is limited by
\begin{equation}
\left\| {p_{2k - 1,2k}^{\rm{c}}} \right\|_2^2 + \left\| {p_{2k - 1}^{\rm{p}}} \right\|_2^2 + \left\| {p_{2k}^{\rm{p}}} \right\|_2^2 \le  {P}_\textup{T},
\end{equation}
where $\!\!\left\| {p_{2k}^{\rm{p}}} \right\|_2^2{\rm{ = }}\!\!\sum\limits_{{m_k} = 1}^{M_k} {\sum\limits_{{n_{2k}} = 1}^{N_i} \!\!\!{\left\| {{{p}^{\rm{p}}_{m_k,{n_{2k}}}}} \right\|_2^2} }$, \!\!\!\!$\left\| {p_{2k - 1}^{\rm{p}}} \right\|_2^2{\rm{ = }}\!\!\!\!\sum\limits_{{m_k} = 1}^{M_k} \!\!{\sum\limits_{{n_{2k \!-\! 1}} = 1}^{N_i}}$
${\left\| {{{  p}^{\rm{p}}_{{m_k},{n_{2k - 1}}}}} \right\|_2^2} $,
$\left\| {p_{2k - 1,2k}^{\rm{c}}} \right\|_2^2 {\rm{ = }}\sum\limits_{{m_k} = 1}^{M_k} {\sum\limits_{{n_k} = 1}^{N_i} {\left\| {{{  p}^c_{{m_k},n_k}}} \right\|_2^2}}$.
$\left\| {\cdot} \right\|_2^2$ is the second-order norm, ${P}_\textup{T}$ denotes the maximum transmit power,
${  p^{\rm{p}}_{m_k,{n_{2k-1}}}}$ represents the precoding power amplitude of the private message transmitted from the $m_k$-th antenna element to the $n_{2k-1}$-th antenna element of ${\rm{U}}{{\rm{D}}_{2k-1}}$,
${  p^{\rm{p}}_{m_k,{n_{2k}}}}$ represents the precoding power amplitude of the private message transmitted from the $m_k$-th antenna element to the $n_{2k}$-th antenna element of ${\rm{U}}{{\rm{D}}_{2k}}$
and ${  p^c_{{m_k},n_k}}$ represents the precoding power amplitude of the common information transmitted from ${m_k}$-th antenna element to $n_k$-th antenna element of all users. The precoding matrix can be expressed as
\begin{equation}
{\emph{\textbf{P}}} = \left[ {\begin{array}{*{20}{c}}&{p_{2k - 1}^{\rm{p}}}&{p_{2k}^{\rm{p}}}&{p_{2k - 1,2k}^{\rm{c}}}\end{array}} \right].
\end{equation}
%where $p_{2k - 1}^{\rm{p}}$, $p_{2k}^{\rm{p}}$, $p_{2k - 1,2k}^{\rm{c}}  \in { \mathbb{C}^{M_k \times N_i}}$.

Multiple OAM modes are used to transmit different RS signals to different user pairs. By utilizing the carrier with different OAM modes, there is no inter-channel interference at the same frequency and time slot.
The RS signal ${x_k}$ is simultaneously transmitted by $M_k$ antenna elements, which is the vortex wave signal transmitted by the carrier with OAM mode ${l_k}$. For the $m_k$-th antenna element, the transmit vortex wave signal is expressed as
\begin{equation}\label{7666}
\!\!\!\!\!\!\!\!\!\!\!\!\!\!\!\!\!\!\!\!\!\!{X _{m_k}} = \sum\limits_{{l_k} \in L} {\left( {{x_k}} \right){e^{j\left( {{\psi _{m_k}} + {\eta _{m_k}}} \right){l_k}}}} \nonumber
\end{equation}
\begin{equation}\label{7}
~~= \sum\limits_{{l_k} \in L} {\left( {{x_k}} \right){e^{j\left( {\frac{{2\pi ({m_k} - 1)}}{M_k} + {\eta_{m_k}}} \right){l_k}}}} ,
\end{equation}
%\begin{equation}
%\!\!\!\!\!\!\!\!\!\! = \left( {{s_k}} \right){e^{j\left( {\frac{{2\pi (n - 1)}}{N} + {\alpha _{{r_n}}}} \right){l_n}}},
%\end{equation}
where ${l_k}$ denotes the OAM mode utilized by the $k$-th user pair $({\rm{U}}{{\rm{D}}_{2k-1}}, {\rm{U}}{{\rm{D}}_{2k}})$, and belongs to the OAM mode set $L = \{ {l_k}|k \in [1,K]\}$. In (\ref{7666}), ${\psi _{m_k}} = 2\pi ({m_k} - 1)/M_k$ represents the basic angle of the transmit UCA antenna element. ${\psi _{m_k}} + {\eta_{m_k}}$ denotes the azimuth angle that is the angular position between the $x$-axis on the plane $z=0$ and $m_k$-th antenna element. Furthermore, ${\eta_{m_k}}$ is the angle between respective antenna element and the phase angle.

The receive vortex wave signal goes through the OAM channel with zero mean circular symmetric complex gaussian (ZMCSCG) noise. By using the SIC technique, ${\rm{U}}{{\rm{D}}_{2k-1}}$ and ${\rm{U}}{{\rm{D}}_{2k}}$ respectively decode their private signals ${\hat x_{2k-1}^{\rm{p}}}$ and ${\hat x_{2k}^{\rm{p}}}$. To obtain ${\hat x_k^{\rm{c}}}$, ${\rm{U}}{{\rm{D}}_{2k-1}}$ and ${\rm{U}}{{\rm{D}}_{2k}}$ will decode the common message. The OAM signals received by ${\rm{U}}{{\rm{D}}_{2k-1}}$ and ${\rm{U}}{{\rm{D}}_{2k}}$ are respectively written as
\begin{small}
\begin{equation}
\begin{array}{l}
{y_{2k-1}} =\\
\!\!\!\! \sum\limits_{{n_{2k-1}} = 1}^{{N_{2k-1}}}\! \!{\left(\!\!\!{\left( {\sum\limits_{{m_k} = 1}^{M_k} {\sum\limits_{{l_k} \in L} \!\!\!{{X _{m_k}}{h_{{m_k},{n_{2k - 1}}}}{e^{j\left( {\frac{{2\pi ({m_k} - 1)}}{M_k} + {\eta _{m_k}}} \right){l_k}}}} } } \!\right) \!\!+ \!{z_{{n_{2k-1}}}}}\!\! \right)}
\end{array}
 \end{equation}
 \end{small}
 and
 \begin{equation}
 \begin{array}{l}
 {y_{2k}} =\\
 \!\!\sum\limits_{{n_{2k}} = 1}^{{N_{2k}}}\!\! {\left(\!\!\! {\left( {\sum\limits_{{m_k} = 1}^{M_k} {\sum\limits_{{l_k} \in L}^{} \!\!\!{{X _{m_k}}{h_{{m_k},{n_{2k}}}}{e^{j\left( {\frac{{2\pi ({m_k} - 1)}}{M_k} +{\eta _{m_k}}} \right){l_k}}}} } } \right)\! + \!{z_{{n_{2k}}}}} \right)},
 \end{array}
 \end{equation}
where $h_{{m_k},{n_{2k - 1}}}$ and $h_{{m_k},{n_{2k - 1}}}$ denote the channel coefficients from the $m_k$-th antenna element of the $k$-th transmit UCA to $n_{2k - 1}$-th antenna element of the ${\rm{U}}{{\rm{D}}_{2k - 1}}$ receive UCA and $n_{2k}$-th antenna element of the ${\rm{U}}{{\rm{D}}_{2k}}$ receive UCA, respectively; ${z_{{n_{2k-1}}}}$ and ${z_{{n_{2k}}}}$ represent the ZMCSCG noise with zero mean and variance $\sigma^2$ received by the $n_{2k-1}$-th antenna element of the ${\rm{U}}{{\rm{D}}_{2k-1}}$ receive UCA and $n_{2k}$-th antenna element of the ${\rm{U}}{{\rm{D}}_{2k}}$ receive UCA, respectively.

\section{Channel Model and Capacity Maximization Problem}
The channel coefficient from the $m_k$-th antenna element of the $k$-th transmit UCA to $n_i$-th antenna element of the ${\rm{U}}{{\rm{D}}_i}$ receive UCA is derived as
\begin{equation}
{h_{{m_k},{n_i}}} = {A_{k,{n_i}}}{e^{ - j({B_{k,{n_i}}}\sin ({\psi _{m_k}} + {\eta_{m_k}} - {\varphi _{{n_i}}} - {\eta _{n_i}} + {\zeta _{k,{n_i}}}))}},
\end{equation}
where
\begin{equation}
\begin{aligned}
{A_{k,{n_i}}}=& \frac{{\beta \lambda }}{{4\pi \sqrt {d_{k,i}^2 + R_k^2 + r_{i}^2} }}{e^{ - \frac{{j2\pi }}{\lambda }\sqrt {d_{k,i}^2 + R_k^2 + r_{i}^2} }}\\
&{e^{j2\pi r_{i}^{}d_{k,i}^{}\sin {\phi _i}\frac{{\cos ({\varphi _{{n_i}}} + {\eta _{n_i}} - \theta )}}{{\lambda \sqrt {d_{k,i}^2 + R_k^2 + r_{i}^2} }}}}
\end{aligned}
\end{equation}
and
\begin{small}
\begin{equation}
\begin{aligned}
&{B_{k,{n_i}}}= \\
&\frac{{2\pi {R_k}\sqrt {r_{i}^2 + d_{k,i}^2{{\sin }^2}{\phi _i} - 2{r_{i}}d_{k,i}^{}\sin {\phi _i}\cos ({\varphi _{{n_i}}} + {\eta _{n_i}} - \theta )} }}{{\lambda \sqrt {d_{k,i}^2 + R_k^2 + r_{i}^2} }}.
\end{aligned}
\end{equation}
\end{small}
For the ${\rm{U}}{{\rm{D}}_i}$ receive UCA, ${\varphi _{{n_i}}} = 2\pi ({n_i} - 1)/{N_i}$ is the basic angle of the ${n_i}$-th antenna element. $\beta $ denotes the phase rotation error and attenuation caused by antenna property. ${R_k}$ is the radius of the $k$-th transmit UCA. ${{r}_{i}}$ represents the radius of the ${\rm{U}}{{\rm{D}}_i}$ receive UCA. $\lambda $ denotes the wavelength. ${\eta _{n_i}}$ denotes the angle between the ${\rm{U}}{{\rm{D}}_i}$ receive UCA and initial phase of transmit UCA. $\theta$ represents the angle between the $x$-axis and the central line projection of the transmit UCA and ${\rm{U}}{{\rm{D}}_i}$ receive UCA. ${\zeta _{k,{n_i}}}$ is defined as
\begin{equation}
\begin{aligned}
&\sin {\zeta _{k,{n_i}}} =\\
 &\frac{{{r_{i}} - d_{k,i}^{}\sin {\phi _i}\cos ({\varphi _{{n_i}}} + {\eta _{n_i}} - \theta )}}{{\sqrt {r_{i}^2{\rm{ + }}d_{k,i}^2{{\sin }^2}{\phi _i} - 2{r_{i}}d_{k,i}^{}\sin {\phi _i}\cos ({\varphi _{{n_i}}} + {\eta _{n_i}} - \theta )} }}
\end{aligned}
\end{equation}
and
\begin{equation}
\begin{aligned}
&\cos {\zeta _{k,{n_i}}} =\\
 &\frac{{d_{k,i}^{}\sin {\phi _i}\cos ({\varphi _{{n_i}}} + {\eta _{n_i}} - \theta )}}{{\sqrt {r_{i}^2{\rm{ + }}d_{k,i}^2{{\sin }^2}{\phi _i} - 2{r_{i}}d_{k,i}^{}\sin {\phi _i}\cos ({\varphi _{{n_i}}} + {\eta _{n_i}} - \theta )} }},
\end{aligned}
\end{equation}
where ${\phi _i}$ is the angle between the $z$-axis and the central connection from the transmit UCA to the ${\rm{U}}{{\rm{D}}_i}$ receive UCA.

For the private message, the signal-to-interference-plus-noise ratios (SINRs) of ${\rm{U}}{{\rm{D}}_{2k-1}}$ and ${\rm{U}}{{\rm{D}}_{2k}}$ are respectively expressed as
\begin{equation}
\begin{array}{l}
\gamma _{{{    m}_k},{{   n}_{2k - 1}}}^{\rm{p}} =\\
 \!\!{\left| {h_{{{    m}_k},{{    n}_{2k - 1}}}^{\rm{T}}  p_{{{    m}_k},{{    n}_{2k - 1}}}^{\rm{p}}} \right|^2}\bigg/(\sum\limits_{{n_{2i}} \in \textbf{\rm{N}}} {\sum\limits_{{m_i} \in \textbf{\rm{M}}} {{{\left| {h_{{{    m}_k},{{     n}_{2k - 1}}}^{\rm{T}}  p_{{m_i},{n_{2i}}}^{\rm{p}}} \right|}^2}} } \\
~~~~~~~~~~~~~ +\!\!\!\!\!\!\! \sum\limits_{\scriptsize{\begin{array}{*{20}{c}}
{{n_{2i - 1}}\!\! \in\!\! \textbf{\rm{N}}}\\
{{n_{2i - 1}}\! \!\notin\!\! {{    n}_{2k - 1}}}
\end{array}}}\!\!\!\!\!\!\!{\sum\limits_{\scriptsize{\begin{array}{*{20}{c}}
{{m_i}\!\! \in \!\!\textbf{\rm{M}}}\\
{{m_i}\!\! \notin \!{{   m}_k}}
\end{array}}}  \!\!\!\!{{{\left| {h_{{{    m}_k},{{    n}_{2k - 1}}}^{\rm{T}}{{  p}_{{m_i},{n_{2i - 1}}}^{\rm{p}}}} \right|}^2}} }  + {\sigma ^2})
\end{array}
\end{equation}
and
 \begin{equation}
\begin{array}{l}
\gamma _{{{    m}_k},{{    n}_{2k}}}^{\rm{p}} =\\
{\left| {h_{{{    m}_k},{{    n}_{2k}}}^{\rm{T}}{{  p}_{{{    m}_k},{{    n}_{2k}}}^{\rm{p}}}} \right|^2}\bigg/(\sum\limits_{{n_{2i - 1}} \in \textbf{\rm{N}}} {\sum\limits_{{m_i} \in \textbf{\rm{M}}} {{{\left| {h_{{{    m}_k},{{    n}_{2k}}}^{\rm{T}}{{  p}_{{m_i},{n_{2i - 1}}}^{\rm{p}}}} \right|}^2}} } \\
~~~~~~~~~~~~~~~~~~~+\!\!\!\!\! \sum\limits_{\scriptsize{\begin{array}{*{20}{c}}
{{n_{2i}}\!\! \in\!\! \textbf{\rm{N}}}\\
{{n_{2i}}\!\! \notin\!\! {{    n}_{2k}}}
\end{array}}} \!\!\!\!\!\!{\sum\limits_{\scriptsize{\begin{array}{*{20}{c}}
{{m_i}\!\! \in\!\! \textbf{\rm{M}}}\\
{{m_i}\!\! \notin \!{{    m}_k}}
\end{array}}} \!\!\!{{{\left| {h_{{{    m}_k},{{    n}_{2k}}}^{\rm{T}}{{  p}_{{m_i},{n_{2i}}}^{\rm{p}}}} \right|}^2}} }  + {\sigma ^2})
\end{array}
 \end{equation}
For the common part, the SINRs of ${\rm{U}}{{\rm{D}}_{2k-1}}$ and ${\rm{U}}{{\rm{D}}_{2k}}$ can be respectively written as
 \begin{equation}
 \begin{array}{l}
\!\!\!\!\gamma _{{{    m}_k},{{   n}_{2k - 1}}}^{\rm{c}} = \\
\!\!\!\!{\left| {h_{{{   m}_k},{{   n}_{2k - 1}}}^{\rm{T}}{{  p}_{{{   m}_k},   n_k}^{\rm{c}}}} \right|^2} \!\bigg/ \! ({\sigma ^2} +\!\!\!\!\! \sum\limits_{{n_{2i - 1}} \in \textbf{\rm{N}}} \!{\sum\limits_{{m_i} \in \textbf{\rm{M}}} \!\!{{{\left| {h_{{{   m}_k},{{   n}_{2k - 1}}}^{\rm{T}}{{  p}_{{m_i},{n_{2i - 1}}}^{\rm{p}}}} \right|}^2}} } \\
\!\!\!\!\!\! + \!\!\!\sum\limits_{{n_{2i}} \in \textbf{\rm{N}}} \!{\sum\limits_{{m_i} \in \textbf{\rm{M}}} \!\!{{{\left| {h_{{{   m}_k},{{   n}_{2k - 1}}}^{\rm{T}}{{  p}}_{{m_i},{n_{2i}}}^{\rm{p}}} \!\right|}^2}}} \!\!\!\! + \!\!\!\!\!\!\!\!\!\!\sum\limits_{\scriptsize{\begin{array}{*{20}{c}}
{ n_i \!\!\in\!\! \textbf{\rm{N}}}\\
{ n_i \!\!\notin\!\!     n_{2k - 1}}
\end{array}}} \!\!\!\!\!\!\!{\sum\limits_{\scriptsize{\begin{array}{*{20}{c}}
{{m_i}\!\! \in \!\!\textbf{\rm{M}}}\\
{{m_i} \!\!\notin\!\! {{   m}_k}}
\end{array}}} \!\!\!\!\!{{{\left| h_{{{   m}_k},{{   n}_{2k - 1}}}^{\rm{T}}{{  p}_{{m_i}, n_i} ^{\rm{c}}} \right|}^2}} }\!\!)
\end{array}
 \end{equation}
 and
 \begin{equation}
\begin{array}{l}
\!\!\!\!\gamma _{{{   m}_k},{{   n}_{2k}}}^{\rm{c}} =\\
\!\!\!\!{\left| {h_{{{   m}_k},{{   n}_{2k}}}^{\rm{T}}{{  p}_{{{   m}_k},    n_k}^{\rm{c}}}} \right|^2}\!\bigg/\!({\sigma ^2} +\!\!\! \sum\limits_{{n_{2i - 1}} \in \textbf{\rm{N}}} {\sum\limits_{{m_i} \in \textbf{\rm{M}}} {{{\left| {h_{{{   m}_k},{{    n}_{2k}}}^{\rm{T}}{{  p}_{{m_i},{n_{2i - 1}}}^{\rm{p}}}} \right|}^2}} } \\
\!\!\!\!\!\!+ \!\!\!\sum\limits_{{n_{2i}} \in \textbf{\rm{N}}} \!{\sum\limits_{{m_i} \in \textbf{\rm{M}}}\!\!\!\! {{{\left| {h_{{{   m}_k},{{   n}_{2k}}}^{\rm{T}}{{  p}_{{m_i},{n_{2i}}}^{\rm{p}}}} \right|}^2}} }
\!\! + \!\!\!\!\!\!\!\!\!\sum\limits_{\scriptsize{\begin{array}{*{20}{c}}
{ n_i \!\!\in \textbf{\rm{N}}}\\
{ n_i \!\notin\!     n_{2k}}
\end{array}}} \!\!\!\!\!\!\!{\sum\limits_{\scriptsize{\begin{array}{*{20}{c}}
{{m_i}\!\! \in \textbf{\rm{M}}}\\
{{m_i}\! \notin \!{{   m}_k}}
\end{array}}}\!\!\!\!\!\!\!\! {{{\left| {h_{{{   m}_k},{{   n}_{2k}}}^{\rm{T}}{{  p}_{{m_i}, n_i}^{\rm{c}}}} \right|}^2}} } )
\end{array}
\end{equation}

The private capacities of ${\rm{U}}{{\rm{D}}_{2k-1}}$ and ${\rm{U}}{{\rm{D}}_{2k}}$ are respectively written as
 \begin{equation}\label{14}
C_{2k - 1}^{\rm{p}} = \!\!\!\sum\limits_{\scriptsize{\begin{array}{*{10}{c}}{{{  n}_{2k - 1}} \!\!\in \!\! \textbf{\rm{N}}}\\
{{{   m}_k}\! \in \!\textbf{\rm{M}}}\end{array}}} \!\!\!\!{\sum\limits_{{q_{2k - 1}} \in {Q_{2k - 1}}} \!\!\!{{{\log }_2}\left( {1 + \frac{{\gamma _{{{   m}_k},{{   n}_{2k - 1}}}^{\rm{p}}}}{M_k}\tau _{{q_{2k - 1}}}^2} \right)} }
 \end{equation}
 and
 \begin{equation}\label{15}
C_{2k}^{\rm{p}} = \!\!\!\sum\limits_{\scriptsize{\begin{array}{*{10}{c}}{{{  n}_{2k}} \!\!\in\!\! \textbf{\rm{N}}}\\
{{{   m}_k} \!\in \!\textbf{\rm{M}}}\end{array}}}\!\!\!\!{\sum\limits_{{q_{2k}} \in {Q_{2k}}} \!\!{{{\log }_2}\left( {1 + \frac{{\gamma _{{{   m}_k},{{   n}_{2k}}}^{\rm{p}}}}{M_k}\tau _{{q_{2k}}}^2} \right)} }.
 \end{equation}
The common capacities of ${\rm{U}}{{\rm{D}}_{2k-1}}$ and ${\rm{U}}{{\rm{D}}_{2k}}$ are respectively derived as
 \begin{equation}\label{16}
C_{2k - 1}^{\rm{c}} = \!\!\!\sum\limits_{\scriptsize{\begin{array}{*{10}{c}}{{{  n}_{2k - 1}}\!\! \in \!\! \textbf{\rm{N}}}\\
{{{    m}_k}\! \in \!\textbf{\rm{M}}}\end{array}}} \!\!\!\!{\sum\limits_{{q_{2k - 1}} \in {Q_{2k - 1}}} \!\!{{{\log }_2}\left( {1 + \frac{{\gamma _{{{   m}_k},{{   n}_{2k - 1}}}^c}}{M_k}\tau _{{q_{2k - 1}}}^2} \right)} }
 \end{equation}
 and
 \begin{equation}\label{17}
C_{2k}^{\rm{c}} = \!\!\!\sum\limits_{\scriptsize{\begin{array}{*{10}{c}}{{{  n}_{2k}}\!\! \in\!\! \textbf{\rm{N}}}\\
{{{    m}_k}\! \in \!\textbf{\rm{M}}}\end{array}}}\!\!\!\!{\sum\limits_{{q_{2k}} \in {Q_{2k}}} \!\!{{{\log }_2}\left( {1 + \frac{{\gamma _{{{   m}_k},{{   n}_{2k}}}^c}}{M_k}\tau _{{q_{2k}}}^2} \right)} },
 \end{equation}
where $\tau _{{q_{2k-1}}}^2$ and $\tau _{{q_{2k}}}^2$ are respectively the eigenvalues of channel matrix from the BS to ${\rm{U}}{{\rm{D}}_{2k-1}}$ and ${\rm{U}}{{\rm{D}}_{2k}}$.

Moreover, the ranks of the channel matrix are ${Q_{2k-1}} \le {R_{2k-1}} = \min ({N_{2k-1}},M_k)$ and ${Q_{2k}} \le {R_{2k}} = \min ({N_{2k}},M_k)$. The downlink OAM-MIMO communications with RS will broadcast the common message to ${\rm{U}}{{\rm{D}}_{2k-1}}$ and ${\rm{U}}{{\rm{D}}_{2k}}$, and thus the common capacity is derived as $C_{2k - 1,2k}^{\rm{c}} = \min \{  C_{2k-1}^{\rm{c}},\; C_{2k}^{\rm{c}}\} $, which guarantees that ${\rm{U}}{{\rm{D}}_{2k-1}}$ and ${\rm{U}}{{\rm{D}}_{2k}}$ can receive the common message. The common capacities of ${\rm{U}}{{\rm{D}}_{2k-1}}$ and ${\rm{U}}{{\rm{D}}_{2k}}$ are respectively ${c_{2k-1}}$ and ${c_{2k}}$, which satisfy the constraint ${c_{2k-1}} + {c_{2k}} \le {C_{2k - 1,2k}^{\rm{c}}}$. In this paper, we consider the full rank channel matrix (i.e., ${Q_{2k-1}} = {R_{2k-1}}$ and ${Q_{2k}} = {R_{2k}}$).
The capacities of ${\rm{U}}{{\rm{D}}_{2k-1}}$ and ${\rm{U}}{{\rm{D}}_{2k}}$ are respectively derived as follows
 \begin{equation}
 C_{2k-1}^{\rm{RS}} = C_{2k-1}^{\rm{p}} + {c_{2k-1}}
 \end{equation}
 and
 \begin{equation}
 C_{2k}^{\rm{RS}} = C_{2k}^{\rm{p}} + {c_{2k}}.
 \end{equation}
The sum capacity of the downlink OAM-MIMO communications with RS is expressed as
 \begin{equation}\label{20}
 C_{\rm{sum}}^{\rm{RS}} = C_{2k-1}^{\rm{p}} + C_{2k}^{\rm{p}} + {c_{2k-1}} + {c_{2k}}.
 \end{equation}
The optimization problem is to maximize the sum capacity $C_{\rm{sum}}^{\rm{RS}}$.

\section{Fractional Programming Based Precoding Optimization Algorithm}

The FP with quadratic transformation is introduced to obtain the optimal precoding matrix for the sum capacity maximization problem.
%(\ref{14}), (\ref{15}), (\ref{16}), (\ref{17})
The private and common capacities of ${\rm{U}}{{\rm{D}}_{2k-1}}$ and ${\rm{U}}{{\rm{D}}_{2k}}$ can be respectively transformed in the following.

The private capacity of ${\rm{U}}{{\rm{D}}_{2k - 1}}$ is derived as
\begin{equation}
\begin{aligned}
\!\!\!\!\!\!C_{2k - 1}^{\rm{p}}& =\!\! \sum\limits_{\scriptsize{\begin{array}{*{20}{c}}
{{  n_{2k - 1}}\!\! \in\!\! \textbf{\rm{N}}}\\
{{  m_k}\! \in\! \textbf{\rm{M}}}
\end{array}}} \!\!\!\! C_{{  m_k},{  n_{2k - 1}}}^{\rm{p}}\\
& =\!\!\!\! \sum\limits_{\scriptsize{\begin{array}{*{20}{c}}
{{  n_{2k - 1}}\!\! \in\!\! \textbf{\rm{N}}}\\
{{  m_k}\! \in\! \textbf{\rm{M}}}
\end{array}}}\!\!\!\! {\sum\limits_{{q_{2k - 1}} \in {Q_{2k - 1}}} \!\!\!\!{{{\log }_2}(1 + \frac{{  \bar \gamma _{{  m_k},{  n_{2k - 1}}}^{\rm{p}}}}{M_k}\tau _{{q_{2k - 1}}}^2)} },
\end{aligned}
\end{equation}
where
\begin{equation}
\bar \gamma _{{  m_k},{  n_{2k - 1}}}^{\rm{p}} = {(a_{{  m_k},{  n_{2k - 1}}}^{\rm{p}})^{\rm{T}}}{(b_{{  m_k},{  n_{2k - 1}}}^{\rm{p}})^{ - 1}}a_{{  m_k},{  n_{2k - 1}}}^{\rm{p}}.
\end{equation}
$a_{{{    m}_k},{{    n}_{2k - 1}}}^{\rm{p}}$ and $b_{{{    m}_k},{{    n}_{2k - 1}}}^{\rm{p}}$ can be respectively defined as
\begin{equation}\label{23}
a_{{{    m}_k},{{    n}_{2k - 1}}}^{\rm{p}} = {h_{{{    m}_k},{{    n}_{2k - 1}}}}  p_{{{    m}_k},{{ n}_{2k - 1}}}^{\rm{p}}
\end{equation}
and
\begin{equation}\label{24}
\begin{array}{*{20}{l}}
~~{b_{{  m_k},{  n_{2k - 1}}}^{\rm{p}}{\rm{ = }}~~h_{{  m_k},{  n_{2k - 1}}}^{}(\sum\limits_{{n_{2i}} \in \textbf{\rm{N}}} {{{\sum\limits_{{m_i} \in \textbf{\rm{M}}} {p_{{m_i},{n_{2i}}}^{\rm{p}}(p_{{m_i},{n_{2i}}}^{\rm{p}})} }^{\rm{T}}}} }\\
{\!\!\! + \!\!\sum\limits_{\scriptsize{\begin{array}{*{20}{c}}
{{n_{2i - 1}}\!\! \in\!\! \textbf{\rm{N}}} \\
{{ n_{2i - 1}}\!\! \notin\!\!   n_{2k - 1}}
\end{array}}}\!\!\!\!\!\!\!\!{\sum\limits_{\scriptsize{\begin{array}{*{20}{c}}
{{m_i}\!\! \in\!\! \textbf{\rm{M}}} \\
{{m_i}\!\! \notin\!\! {  m_k}}
\end{array}}}\!\!\!\! p_{{m_i},{n_{2i - 1}}}^{\rm{p}}{{(p_{{m_i},{n_{2i - 1}}}^{\rm{p}})}^{\rm{T}}}} )h_{{  m_k},{  n_{2k - 1}}}^{\rm{T}} \!\!+\! {\sigma ^2}}.
\end{array}
\end{equation}
The private capacity of ${\rm{U}}{{\rm{D}}_{2k}}$ is expressed as
\begin{equation}
\begin{aligned}
\!\!\!\!C_{2k}^{\rm{p}}& = \!\!\sum\limits_{\scriptsize{\begin{array}{*{20}{c}}
{{  n_{2k}}\!\! \in\!\! \textbf{\rm{N}}}\\
{{  m_k}\! \in \!\textbf{\rm{M}}}
\end{array}}} \!\!\!\! C_{{  m_k},{  n_{2k}}}^{\rm{p}}\\
~~ & =\!\!\!\! \sum\limits_{\scriptsize{\begin{array}{*{20}{c}}
{{  n_{2k}}\!\! \in \!\!\textbf{\rm{N}}}\\
{{  m_k} \!\in \!\textbf{\rm{M}}}
\end{array}}} \!\!\!\!{\sum\limits_{{q_{2k}} \in {Q_{2k}}} {{{\log }_2}(1 + \frac{{  \bar \gamma _{{  m_k},{  n_{2k}}}^{\rm{p}}}}{M_k}\tau _{{q_{2k}}}^2)} },
\end{aligned}
\end{equation}
where
\begin{equation}
\bar \gamma _{{  m_k},{  n_{2k}}}^{\rm{p}} = {(a_{{  m_k},{  n_{2k}}}^{\rm{p}})^{\rm{T}}}{(b_{{  m_k},{  n_{2k}}}^{\rm{p}})^{ - 1}}a_{{  m_k},{  n_{2k}}}^{\rm{p}}.
\end{equation}
$a_{{{   m}_k},{{   n}_{2k}}}^{\rm{p}}$ and $b_{{{   m}_k},{{   n}_{2k}}}^{\rm{p}}$ can be respectively defined as
\begin{equation}\label{27}
a_{{{   m}_k},{{   n}_{2k}}}^{\rm{p}} = h_{{{   m}_k},{{   n}_{2k}}}^{}  p_{{{   m}_k},{{   n}_{2k}}}^{\rm{p}}
\end{equation}
and
\begin{equation}\label{28}
\begin{array}{*{20}{l}}
{b_{{  m_k},{  n_{2k}}}^{\rm{p}} = h_{{  m_k},{  n_{2k}}}^{}(\sum\limits_{{n_{2i - 1}} \in \textbf{\rm{N}}} {{{\sum\limits_{{m_i} \in \textbf{\rm{M}}} {p_{{m_i},{n_{2i - 1}}}^{\rm{p}}(p_{{m_i},{n_{2i - 1}}}^{\rm{p}})} }^{\rm{T}}}} }\\
~~~~~~~~~~{ + \!\!\!\!\!\!\sum\limits_{\scriptsize{\begin{array}{*{20}{c}}
{{n_{2i }}\!\! \in\!\! \textbf{\rm{N}}} \\
{{ n_{2i }}\!\! \notin\!\!   n_{2k }}
\end{array}}}\!\!\!\!\!\!\!{\sum\limits_{\scriptsize{\begin{array}{*{20}{c}}
{{m_i} \!\!\in\!\! \textbf{\rm{M}}} \\
{{m_i}\!\! \notin\!\! {  m_k}}
\end{array}}}\!\!\!\! {p_{{m_i},{n_{2i}}}^{\rm{p}}{{(p_{{m_i},{n_{2i}}}^{\rm{p}})}^{\rm{T}}}} } )h_{{  m_k},{  n_{2k}}}^{\rm{T}} + {\sigma ^2}}.
\end{array}
\end{equation}

The common capacity of ${\rm{U}}{{\rm{D}}_{2k - 1}}$ is derived as
\begin{equation}
\begin{aligned}
\!\!\!\!C_{2k - 1}^{\rm{c}} &= \!\!\sum\limits_{\scriptsize{\begin{array}{*{20}{c}}
{{  n_{2k - 1}} \!\!\in \!\!\textbf{\rm{N}}}\\
{{  m_k}\! \in\! \textbf{\rm{M}}}
\end{array}}} \!\!\!\! C_{{  m_k},{  n_{2k - 1}}}^{\rm{c}}\\
 ~~~~~&= \!\!\!\!\sum\limits_{\scriptsize{\begin{array}{*{20}{c}}
{{  n_{2k - 1}} \!\!\in\!\! \textbf{\rm{N}}}\\
{{  m_k}\! \in\! \textbf{\rm{M}}}
\end{array}}}\!\!\!\! {\sum\limits_{{q_{2k - 1}} \in {Q_{2k - 1}}} {{{\log }_2}(1 + \frac{{ \bar \gamma _{{  m_k},{  n_{2k - 1}}}^{\rm{c}}}}{M_k}\tau _{{q_{2k - 1}}}^2)} },
\end{aligned}
\end{equation}
where
\begin{equation}
\bar \gamma _{{  m_k},{  n_{2k - 1}}}^{\rm{c}} = {(a_{{  m_k},{  n_{2k - 1}}}^{\rm{c}})^{\rm{T}}}{(b_{{  m_k},{  n_{2k - 1}}}^{\rm{c}})^{ - 1}}a_{{  m_k},{  n_{2k - 1}}}^{\rm{c}}.
\end{equation}
$a_{{{    m}_k},{{    n}_{2k - 1}}}^{\rm{c}}$ and $b_{{{    m}_k},{{    n}_{2k - 1}}}^{\rm{c}}$ can be respectively defined as
\begin{equation}\label{31}
a_{{{    m}_k},{{    n}_{2k - 1}}}^{\rm{c}} = h_{{{    m}_k},{{    n}_{2k - 1}}}^{}{  p_{{{    m}_k},{{ n_k}}}^{\rm{c}}}
\end{equation}
and
\begin{equation}\label{32}
\begin{aligned}
\!\!\!\!{b_{{  m_k},{  n_{2k - 1}}}^{\rm{c}}\!\! ={\sigma ^2}\!+\! h_{{  m_k},{  n_{2k - 1}}}^{}\!(\!\sum\limits_{{n_{2i - 1}} \in \textbf{\rm{N}}} \!{{{\sum\limits_{{m_i} \in \textbf{\rm{M}}} \!\!{p_{{m_i},{n_{2i}}}^{\rm{p}}(p_{{m_i},{n_{2i}}}^{\rm{p}})} }^{\rm{T}}}} }\\
 + \sum\limits_{{n_{2i - 1}} \in \textbf{\rm{N}}} {{{\sum\limits_{{m_i} \in \textbf{\rm{M}}} {p_{{m_i},{n_{2i - 1}}}^{\rm{p}}(p_{{m_i},{n_{2i - 1}}}^{\rm{p}})} }^{\rm{T}}}}\\
+\!\!\!\! \sum\limits_{\scriptsize{\begin{array}{*{20}{c}}
{n_i \!\!\in\!\! \textbf{\rm{N}}}\\
{n_i \!\!\notin\!\! {  n_{2k - 1}}}
\end{array}}}\!\!\!\!\!\!\!{\sum\limits_{\scriptsize{\begin{array}{*{20}{c}}
{{m_i}\!\! \in\!\! \textbf{\rm{M}}} \\
{{m_i}\!\! \notin\!\! {  m_k}}
\end{array}}}\!\!\!\! {p_{{m_i},n_i}^{\rm{c}}{{(p_{{m_i},n_i}^{\rm{c}})}^{\rm{T}}}} )h_{{  m_k},{  n_{2k - 1}}}^{\rm{T}}} .
\end{aligned}
\end{equation}
The common capacity of ${\rm{U}}{{\rm{D}}_{2k}}$ can be derived as
\begin{equation}
\begin{aligned}
\!\!\!\!C_{2k}^{\rm{c}}& = \!\!\!\sum\limits_{\scriptsize{\begin{array}{*{20}{c}}
{{  n_{2k}}\!\! \in\!\! \textbf{\rm{N}}}\\
{{  m_k} \!\in\! \textbf{\rm{M}}}
\end{array}}} \!\!\!\! C_{{  m_k},{  n_{2k}}}^{\rm{c}}\\
~~ &=\!\!\!\! \sum\limits_{\scriptsize{\begin{array}{*{20}{c}}
{{  n_{2k}} \!\!\in\!\! \textbf{\rm{N}}}\\
{{  m_k}\!\! \in\!\! \textbf{\rm{M}}}
\end{array}}}\!\!\! {\sum\limits_{{q_{2k}} \in {Q_{2k}}} {{{\log }_2}(1 + \frac{{\bar \gamma _{{  m_k},{  n_{2k}}}^{\rm{c}}}}{M_k}\tau _{{q_{2k}}}^2)} },
\end{aligned}
\end{equation}
where
\begin{equation}
\bar \gamma _{{  m_k},{  n_{2k}}}^{\rm{c}} = {(a_{{  m_k},{  n_{2k}}}^{\rm{c}})^{\rm{T}}}{(b_{{  m_k},{  n_{2k}}}^{\rm{c}})^{ - 1}}a_{{  m_k},{  n_{2k}}}^{\rm{c}}.
\end{equation}
$a_{{{    m}_k},{{    n}_{2k}}}^{\rm{c}}$ and $b_{{{    m}_k},{{    n}_{2k}}}^{\rm{c}}$ are respectively defined as
\begin{equation}\label{35}
a_{{  m_k},{  n_{2k}}}^{\rm{c}} = h_{{  m_k},{  n_{2k}}}^{}p_{{  m_k},{  n_k}}^{\rm{c}}
\end{equation}
and
\begin{equation}\label{36}
\begin{aligned}
\!\!\!{b_{{  m_k},{  n_{2k}}}^{\rm{c}} = {\sigma ^2}\!+ \! h_{{  m_k},{  n_{2k}}}\!(\!\sum\limits_{{n_{2i - 1}} \in {\textbf{\rm{N}}}} {{{\sum\limits_{{m_i} \in \textbf{\rm{M}}}\!\! {p_{{m_i},{n_{2i - 1}}}^{\rm{p}}(p_{{m_i},{n_{2i - 1}}}^{\rm{p}})} }^{\rm{T}}}} }\\
~~~~~~~~~~~~~~~~~~~~~~~~~~~~~~{ + \sum\limits_{{n_{2i}} \in \textbf{\rm{N}}} {{{\sum\limits_{{m_i} \in \textbf{\rm{M}}} {p_{{m_i},{n_{2i}}}^{\rm{p}}(p_{{m_i},{n_{2i}}}^{\rm{p}})} }^{\rm{T}}}}}\\
~~~~~~~~~~~~~~~~{+ \!\!\!\!\!\sum\limits_{\scriptsize{\begin{array}{*{20}{c}}
{n_i\!\! \in\!\! \textbf{\rm{N}}}\\
{n_i \!\!\notin\!\! {  n_{2k - 1}}}
\end{array}}}\!\!\!\!\!\!\!{\sum\limits_{\scriptsize{\begin{array}{*{20}{c}}
{{m_i}\!\! \in \!\!\textbf{\rm{M}}} \\
{{m_i}\!\! \notin\!\! {  m_k}}
\end{array}}}\!\!\!\! {p_{{m_i},n_i}^{\rm{c}}{{(p_{{m_i},n_i}^{\rm{c}})}^{\rm{T}}}} )h_{{  m_k},{  n_{2k - 1}}}^{\rm{T}}}}.
\end{aligned}
\end{equation}

Then, the sum capacity maximization problem can be rewritten as problem \textbf{P1}.
\begin{equation}
\!\!\!\!\!\!\!\!\!\!\!\!\!\!\!\!\!\!\!\!\!\!\!\!\!\!\!\!\!\!\!\!\!\!\!\!\!\!\!\!\!\!\!\!\!\!\!\!\!\!\!\!\!\!\!\!\!\!\!\!\!\!\!\!\!\!\!\!\!\!\!\!\!\!\!\!\!\!\!\!\!\!\!\!\!\!\!\!\!\!\!\!\!\!\!\!\!{{\rm{\textbf{P1}:}}}\mathop {{\textbf{\rm{maxmize~}}}}\limits_{\rm{P}} C_{{\rm{sum}}}^{{\rm{RS}}}\nonumber
\end{equation}
\begin{equation}
\begin{array}{l}
~\!\!\!\!\!\!\textbf{{\rm{s}}{\rm{.t}}{\rm{.C1: }}}C_{2k - 1}^{\rm{p}} \le
\!\!\!\!\!\!\!\!\!\!\sum\limits_{\scriptsize{\begin{array}{*{20}{c}}
{{  n_{2k - 1}} \!\!\in \!\!\textbf{\rm{N}}}\\
{{  m_k}\!\! \in\!\! \textbf{\rm{M}}}
\end{array}}}\!\!\!\! {\sum\limits_{{q_{2k - 1}} \in {Q_{2k - 1}}} \!\!\!\!{{{\log }_2}(1 + \frac{{\gamma _{{  m_k},{  n_{2k - 1}}}^{\rm{p}}}}{M_k}\tau _{{q_{2k - 1}}}^2)} } \nonumber
\end{array}
\end{equation}
\begin{equation}
\begin{array}{l}
~\!\!\!\!\!\!\!\!\!\!\!\!\!\!\!\!\!\!\!\!\!\textbf{{\rm{C2:}}}~C_{2k}^{\rm{p}} \le
\!\!\!\sum\limits_{\scriptsize{\begin{array}{*{20}{c}}
{{  n_{2k}}\!\! \in \!\!\textbf{\rm{N}}}\\
{{  m_k}\!\! \in\!\! \textbf{\rm{M}}}
\end{array}}} \!\!\!\!{\sum\limits_{{q_{2k}} \in {Q_{2k}}} {{{\log }_2}(1 + \frac{{\gamma _{{  m_k},{  n_{2k}}}^{\rm{p}}}}{M_k}\tau _{{q_{2k}}}^2)} } \nonumber
\end{array}
\end{equation}
\begin{equation}
\begin{array}{l}
~~~\!\textbf{{\rm{C3:}}}~C_{2k - 1,2k}^{\rm{c}} \le
\!\!\!\!\!\!\sum\limits_{\scriptsize{\begin{array}{*{20}{c}}
{{  n_{2k - 1}}\!\! \in\!\! \textbf{\rm{N}}}\\
{{  m_k} \!\!\in \!\!\textbf{\rm{M}}}
\end{array}}}\!\!\!\! {\sum\limits_{{q_{2k - 1}} \in {Q_{2k - 1}}}\!\!\! \!\!\!\!{{{\log }_2}(1 \!+\! \frac{{\gamma _{{  m_k},{  n_{2k - 1}}}^{\rm{c}}}}{M_k}\tau _{{q_{2k - 1}}}^2\!)} } \nonumber
\end{array}
\end{equation}
\begin{equation}
\begin{array}{l}
~~\!\!\!\!\!\!\!\!\!\textbf{{\rm{C4:}}}~C_{2k - 1,2k}^{\rm{c}} \le
 \!\!\!\sum\limits_{\scriptsize{\begin{array}{*{20}{c}}
{{  n_{2k}}\!\! \in\!\! \textbf{\rm{N}}}\\
{{  m_k}\!\! \in\!\! \textbf{\rm{M}}}
\end{array}}}\!\!\! {\sum\limits_{{q_{2k}} \in {Q_{2k}}} {{{\log }_2}(1 + \frac{{\gamma _{{  m_k},{  n_{2k}}}^{\rm{c}}}}{M_k}\tau _{{q_{2k}}}^2)} }\nonumber
\end{array}
\end{equation}
\begin{equation}
\begin{array}{l}
~~\textbf{{\rm{C5:}}} \sum\limits_{{n_{2k - 1}} \in \textbf{\rm{N}}}\sum\limits_{{m_k} \in \textbf{\rm{M}}}  \left\| {p_{{m_k},{n_{2k - 1}}}^{\rm{p}}} \right\|_2^2 \!\! + \!\!\! \sum\limits_{{n_{2k}} \in \textbf{\rm{N}}} \sum\limits_{{m_k} \in \textbf{\rm{M}}}\!  \left\| {p_{{m_k},{n_{2k}}}^{\rm{p}}} \right\|_2^2\\
~~~~~~~~~~~~~~~~~~~~~~~~~~~~~~~~+ {\sum\limits_{{n_k} \in \textbf{\rm{N}}}\sum\limits_{{m_k} \in \textbf{\rm{M}}} {\left\| {p_{{m_k},{n_k}}^{\rm{c}}} \right\|_2^2} }  \le  {{P_{\rm{T}}}} \nonumber
\end{array}
\end{equation}
where C1 and C2 are the constraints on the private capacities of  ${\rm{U}}{{\rm{D}}_{2k - 1}}$ and ${\rm{U}}{{\rm{D}}_{2k}}$, respectively; C3 and C4 are the constraints on the common capacities of ${\rm{U}}{{\rm{D}}_{2k - 1}}$ and ${\rm{U}}{{\rm{D}}_{2k}}$, respectively; C5 is the total transmit power constraint.

The precoding matrix ${\emph{\textbf{P}}}$ is a non-empty matrix. According to the FP in \cite{46}, problem \textbf{P1} can be converted to the convex problem \textbf{P2}.
\begin{equation}
\!\!\!\!\!\!\!\!\!\!\!\!\!\!\!\!\!\!\!\!\!\!\!\!\!\!\!\!\!\!\!\!\!\!\!\!\!\!\!\!\!\!\!\!\!\!\!\!\!\!\!\!\!\!\!\!\!\!\!\!\!\!\!\!\!\!\!\!\!\!\!\!\!\!\!\!\!\!\!\!\!\!\!\!\!\!\textbf{{\rm{\textbf{P2:}}}}\mathop {\textbf{{\rm{maxmize~}}}}\limits_{\rm{P}}   \bar  C_{{\rm{sum}}}^{{\rm{RS}}}\qquad\nonumber
\end{equation}
\begin{equation}
\begin{array}{l}
\!\!\textbf{{\rm{s}}{\rm{.t.}}~{\rm{C1:}}} C_{2k - 1}^{\rm{p}} \le \\ \!\!\!\!\!\!\sum\limits_{\tiny{\begin{array}{*{10}{c}}{{{  n}_{2k-1}} \!\!\in \!\! \textbf{\rm{N}}}\\
{{{   m}_k} \!\!\in\!\! \textbf{\rm{M}}}\end{array}}} \!\!\!\!\!\!{\sum\limits_{\tiny{\begin{array}{*{10}{c}}{{q_{2k - 1}} \!\! \in \!\! {Q_{2k - 1}}}\end{array}}} \!\!\!\!\!\!\!\!\!\!\!{{{\log }_2}\!\left( \!\!{1 \!+\! \frac{{\tau _{{q_{2k - 1}}}^2}}{M_k}\!\left( \!{2{\mathop{\rm Re}\nolimits} \left\{ {{{\!\left( {y_{{{    m}_k},{{    n}_{2k - 1}}}^{\rm{p}}} \right)}^{\rm{T}}}\!\!\!a_{{{    m}_k},{{    n}_{2k - 1}}}^{\rm{p}}} \!\right\}} \right.} \right.} } \\
~~~~~~~~~~~~~~~~~~~~~~~~\left. {\left. { - {{\left( {y_{{{    m}_k},{{    n}_{2k}}}^{\rm{p}}} \right)}^{\rm{T}}}b_{{{    m}_k},{{    n}_{2k - 1}}}^{\rm{p}}\left( {y_{{{    m}_k},{{    n}_{2k - 1}}}^{\rm{p}}} \right)} \right)} \right)
\end{array}\nonumber
\end{equation}
\begin{equation}
\begin{array}{l}
~~\!~~\textbf{{\rm{C2:}}}C_{2k}^{\rm{p}} \le \\ ~~~~\!\!\!\!\!\sum\limits_{\scriptsize{\begin{array}{*{10}{c}}{{{  n}_{2k}}\!\! \in\!\! \textbf{\rm{N}}}\\
{{{    m}_k}\!\! \in \!\!\textbf{\rm{M}}}\end{array}}}\!\!\!{\sum\limits_{{q_{2k}} \in {Q_{2k}}}\!\!\! {{{\log }_2}\left( {1 + \frac{{\tau _{{q_{2k}}}^2}}{M_k}\left( {2{\mathop{\rm Re}\nolimits} \left\{ {{{\left( {y_{{{    m}_k},{{    n}_{2k}}}^{\rm{p}}} \right)}^{\rm{T}}}a_{{{    m}_k},{{    n}_{2k}}}^p} \right\}} \right.} \right.} } \\
~~~~~~~~~~~~~~~~~~~~~~~~~~~~\left. {\left. { - {{\left( {y_{{{    m}_k},{{    n}_{2k}}}^{\rm{p}}} \right)}^{\rm{T}}}b_{{{    m}_k},{{    n}_{2k}}}^{\rm{p}}\left( {y_{{{    m}_k},{{    n}_{2k}}}^{\rm{p}}} \right)} \right)} \right)
\end{array}\nonumber
\end{equation}
\begin{equation}
\begin{array}{l}
~~\!~~\textbf{{\rm{C3:}}}~C_{2k - 1,2k}^{\rm{c}} \le \\ \!\!\!\!\!\!\sum\limits_{\tiny{\begin{array}{*{10}{c}}{{{  n}_{2k-1}} \!\!\in \!\! \textbf{\rm{N}}}\\
{{{    m}_k} \!\!\in\!\! \textbf{\rm{M}}}\end{array}}} \!\!\!\!\!\!{\sum\limits_{\tiny{\begin{array}{*{10}{c}}{{q_{2k - 1}} \!\! \in \!\! {Q_{2k - 1}}}\end{array}}} \!\!\!\!\!\! \!\!\!\!\!{{{\log }_2}\!\left( \!{1 \!+\! \frac{{\tau _{{q_{2k - 1}}}^2}}{M_k}\!\left( \!\!{2{\mathop{\rm Re}\nolimits} \!\left\{ {{{\!\left( {y_{{{    m}_k},{{    n}_{2k - 1}}}^{\rm{c}}} \right)}^{\rm{T}}}\!\!\!a_{{{    m}_k},{{    n}_{2k - 1}}}^{\rm{c}}} \right\}} \right.} \right.} } \\
~~~~~~~~~~~~~~~~~~~~\left. {\left. { - {{\left( {y_{{{    m}_k},{{    n}_{2k - 1}}}^{\rm{c}}} \right)}^{\rm{T}}}b_{{{    m}_k},{{    n}_{2k - 1}}}^{\rm{c}}\left( {y_{{{    m}_k},{{    n}_{2k - 1}}}^{\rm{c}}} \right)} \right)} \right)\qquad
\end{array}\nonumber
\end{equation}
\begin{equation}
\begin{array}{l}
~~\!~~\textbf{{\rm{C4:}}}~C_{2k - 1,2k}^{\rm{c}} \le \\ ~~~\!\!\!\sum\limits_{\scriptsize{\begin{array}{*{10}{c}}{{{  n}_{2k}}\!\! \in\! \!\textbf{\rm{N}}}\\
{{{    m}_k}\!\! \in\! \!\textbf{\rm{M}}}\end{array}}} \!\!\!{\sum\limits_{{q_{2k}} \in {Q_{2k}}} \!\!\!{{{\log }_2}\left( {1 + \frac{{\tau _{{q_{2k}}}^2}}{M_k}\left( {2{\mathop{\rm Re}\nolimits} \left\{ {{{\left( {y_{{{    m}_k},{{    n}_{2k}}}^{\rm{c}}} \right)}^{\rm{T}}}a_{{{    m}_k},{{    n}_{2k}}}^{\rm{c}}} \right\}} \right.} \right.} } \\
~~~~~~~~~~~~~~~~~~~~~~~~~~~~\left. {\left. { - {{\left( {y_{{{    m}_k},{{    n}_{2k}}}^{\rm{c}}} \right)}^{\rm{T}}}b_{{{    m}_k},{{    n}_{2k}}}^{\rm{c}}\left( {y_{{{    m}_k},{{    n}_{2k}}}^{\rm{c}}} \right)} \right)} \right)
\end{array}\nonumber
\end{equation}
\begin{equation}
\begin{array}{l}
~~\!~~\textbf{{\rm{C5:}}}\!\! \sum\limits_{{n_{2k - 1}} \in \textbf{\rm{N}}} \sum\limits_{{m_k} \in \textbf{\rm{M}}} \! \left\| {p_{{m_k},{n_{2k - 1}}}^{\rm{p}}} \right\|_2^2 \!\! + \!\!\! \sum\limits_{{n_{2k}} \in \textbf{\rm{N}}} \sum\limits_{{m_k} \in \textbf{\rm{M}}} \left\| {p_{{m_k},{n_{2k}}}^{\rm{p}}} \right\|_2^2\\
~~~~~~~~~~~~~~~~~~~~~~~~~~~~~~~~+ \sum\limits_{{n_k} \in \textbf{\rm{N}}} \sum\limits_{{m_k} \in \textbf{\rm{M}}} {\left\| {p_{{m_k},{n_k}}^{\rm{c}}} \right\|_2^2}   \le {{P_{\rm{T}}}} \nonumber
\end{array}
\end{equation}
\begin{equation}
~~~\!\!\!\!\!\!\!\!\!\!\!\!\!\!\!\!\!\!\!\!\!\!\!\!\!\!\!\!\! \textbf{{\rm{C6:}}  } \bar C_{{\rm{sum}}}^{{\rm{RS}}} \le C_{2k - 1}^{\rm{p}} + C_{2k}^{\rm{p}} + C_{2k - 1,2k}^{\rm{c}}\qquad \nonumber
\end{equation}
where $\bar C_{{\rm{sum}}}^{{\rm{RS}}}$ is the sum capacity $C_{{\rm{sum}}}^{{\rm{RS}}}$ transformed from FP optimization algorithm, and the auxiliary variables $y_{{{    m}_k},{{    n}_{2k - 1}}}^{\rm{p}}$, $y_{{{    m}_k},{{    n}_{2k}}}^{\rm{p}}$ and $y_{{{    m}_k},{{    n}_{2k - 1}}}^{\rm{c}}$, $y_{{{    m}_k},{{    n}_{2k}}}^{\rm{c}}$ can be written as
\begin{equation}\label{41}
~~~\!y_{{{    m}_k},{{    n}_{2k - 1}}}^{\rm{p}} = {\left( {b{{_{{{    m}_k},{{    n}_{2k - 1}}}^{\rm{p*}}}}} \right)^{ - 1}}a{_{{{    m}_k},{{    n}_{2k - 1}}}^{\rm{p*}}}
\end{equation}
\begin{equation}\label{42}
y_{{{    m}_k},{{    n}_{2k}}}^{\rm{p}} = {\left( {b{{_{{{    m}_k},{{    n}_{2k}}}^{\rm{p*}}}}} \right)^{ - 1}}a{_{{{    m}_k},{{    n}_{2k}}}^{\rm{p*}}}
\end{equation}
\begin{equation}\label{43}
~~~~\!y_{{{    m}_k},{{    n}_{2k - 1}}}^{\rm{c}} = {\left( {b{{_{{{    m}_k},{{    n}_{2k - 1}}}^{\rm{c*}}}}} \right)^{ - 1}}a{_{{{    m}_k},{{    n}_{2k - 1}}}^{\rm{c*}}}
\end{equation}
\begin{equation}\label{44}
~\!y_{{{    m}_k},{{    n}_{2k}}}^{\rm{c}} = {\left( {b{{_{{{    m}_k},{{    n}_{2k}}}^{\rm{c*}}}}} \right)^{ - 1}}a{_{{{    m}_k},{{    n}_{2k}}}^{\rm{c*}}}
\end{equation}
The iterations of $b{_{{{    m}_k},{{    n}_{2k - 1}}}^{\rm{p*}}}$, $b{_{{{    m}_k},{{    n}_{2k}}}^{\rm{p*}}}$, $b{_{{{    m}_k},{{    n}_{2k - 1}}}^{\rm{c*}}}$, $b{_{{{    m}_k},{{    n}_{2k}}}^{\rm{c*}}}$ and $a{_{{{    m}_k},{{    n}_{2k- 1}}}^{\rm{p*}}}$, $a{_{{{    m}_k},{{    n}_{2k}}}^{\rm{p*}}}$, $a{_{{{    m}_k},{{    n}_{2k - 1}}}^{\rm{c*}}}$, $a{_{{{    m}_k},{{    n}_{2k}}}^{\rm{c*}}}$ are the auxiliary variables calculated by substituting the precoding matrix obtained from the previous iteration into (\ref{24}), (\ref{28}), (\ref{32}), (\ref{36}) and (\ref{23}), (\ref{27}), (\ref{31}), (\ref{35}), respectively. Then, we calculate $  \bar C{_{{\rm{sum}}}^{{\rm{RS*}}}}$ at each iteration through iterative convex approximation. The iteration stops when $\bar C{_{{\rm{sum}}}^{{\rm{RS}}}}$ converges to the convergence threshold $\Gamma$. By substituting the optimal precoding matrix into (\ref{20}), we can obtain the maximum sum capacity $C_{{\rm{sum}}}^{{\rm{RS}}}$. The FP based precoding optimization algorithm is summarized as follows.
\par\noindent\rule[0.25\baselineskip]{8.9cm}{0.5pt}
\uline{\textbf{FP based precoding optimization algorithm}}
\par 1) Initialize ${\emph{\textbf{P}}}^{\rm{*}}$, $\bar C{_{{\rm{sum}}}^{{\rm{RS*}}}}$;
\par 2) Repeat steps 3) \!-\! 7);
\par 3) Update (\ref{41}), (\ref{42}), (\ref{43}), (\ref{44}) to obtain $y_{{{   m}_k},{{   n}_{2k-1}}}^{\rm{p}}$,\par~~~ $y_{{{   m}_k},{{   n}_{2k}}}^{\rm{p}}$,
$y_{{{   m}_k},{{   n}_{2k-1}}}^{\rm{c}}$ and $y_{{{   m}_k},{{   n}_{2k}}}^{\rm{c}}$;
\par 4) By taking the fixed $y_{{{   m}_k},{{   n}_{2k-1}}}^{\rm{p}}$, $y_{{{   m}_k},{{   n}_{2k}}}^{\rm{p}}$, $y_{{{   m}_k},{{   n}_{2k-1}}}^{\rm{c}}$, \par~~~$y_{{{   m}_k},{{   n}_{2k}}}^{\rm{c}}$
and ${\emph{\textbf{P}}}^{\rm{*}}$ into the problem \textbf{P2}, ${\emph{\textbf{P}}}$ and $\bar C_{{\rm{sum}}}^{{\rm{RS}}}$ are\par~~ obtained
by convex approximation;
\par 5) Set ${\emph{\textbf{P}}} = {\emph{\textbf{P}}}^{\rm{*}}$;
\par 6) When $ \bar C_{{\rm{sum}}}^{{\rm{RS}}} -\bar C{_{{\rm{sum}}}^{{\rm{RS*}}}} \leq \Gamma$, go to step 8);
\par 7) Set $\bar C_{{\rm{sum}}}^{{\rm{RS*}}} = \bar C{_{{\rm{sum}}}^{{\rm{RS}}}}$, go to step 3);
\par 8) Substitute the current ${\emph{\textbf{P}}}$ into the problem \textbf{P1} to obtain
\par~~ the maximum sum capacity $C_{{\rm{sum}}}^{{\rm{RS}}}$.
\par\noindent\rule[0.25\baselineskip]{8.9cm}{0.5pt}

\section{Simulation Results}

For the downlink OAM-MIMO communications with RS, the FP based precoding optimization algorithm is demonstrated in simulation results.
%Parameters are set for Fig. 1$ - $3.
%We set $\mathit{UD}_1$ power, $\mathit{UD}_2$ power and sum power as ${p_1} = 0.2$, ${p_2} = 0.8$ and $P = 1$, respectively.
The transmit UCA consists of $3$ antenna elements. The antenna element number of ${\rm{U}}{\rm{D}}$ receive UCA is set to be 4.
The attenuation and phase rotation error caused by antenna property is $\beta=4\pi$.
The radius of the transmit UCA is ${R_k} = k\lambda $ and the radius of the user UCA is ${r_{i}} = n\lambda $.
The wavelength is set to be $\lambda=0.01$.
%The distance from the transmit UCA center to the ${\rm{U}}{{\rm{D}}_{2k-1}}$ and ${\rm{U}}{{\rm{D}}_{2k}}$ receive UCA centers are $10\lambda $ and $20\lambda $, respectively.
The angle between the $x$-axis and the central line projection from the transmit UCA to the receive UCA is $\theta  = 0$. The angle between the $z$-axis and the central connection from the transmit UCA to the receive UCA is set to be ${\phi _i} = 0$.
%The angle between the initial phase of the transmit UCA and the receive UCA ${\alpha _{{r_{mi}}}}$, and the angle between the phase angle and respective array-element ${\alpha _{{r_m}}}$  are zero.
Table \ref{table1} presents the four cases of mode combinations for the downlink OAM-MIMO communications with RS. There are $4$, $3$, $3$ and $2$ antenna elements in transmit UCAs, respectively. The antenna element numbers of ${\rm{U}}{\rm{D}}$ receive UCAs are respectively set to be $4$, $4$, $5$ and $4$.
\begin{table}[htbp]
	\centering
\caption{the cases of different OAM mode combination}
	\label{table1}
\setlength{\tabcolsep}{5.5mm}{
	\begin{tabular}{|c|c|c|c|}
		\hline
		& & & \\[-6pt]
		Case      &   Mode ${l_k}$   & $ N \times M $ &  Eigenvalue ${\tau ^2}$ \\
		\hline
		& & & \\[-6pt]
		1         &   1, 2           & $4\times2$     &   4, 4        \\
        \hline
        & & & \\[-6pt]
	    2         &   1, 2, 3        & $5\times3$     &   5, 5, 5     \\
        \hline
        & & & \\[-6pt]
		3         &   1, 2, 3        & $4\times3$     &  4, 4, 4      \\
        \hline
        & & & \\[-6pt]
		4         &   1, 2, 3, 4     & $4\times4$     &   4, 4, 4, 4  \\
		\hline
	\end{tabular}}

\end{table}

\begin{figure}[thp]
\centering
\includegraphics[height=2.8in,width=9cm]{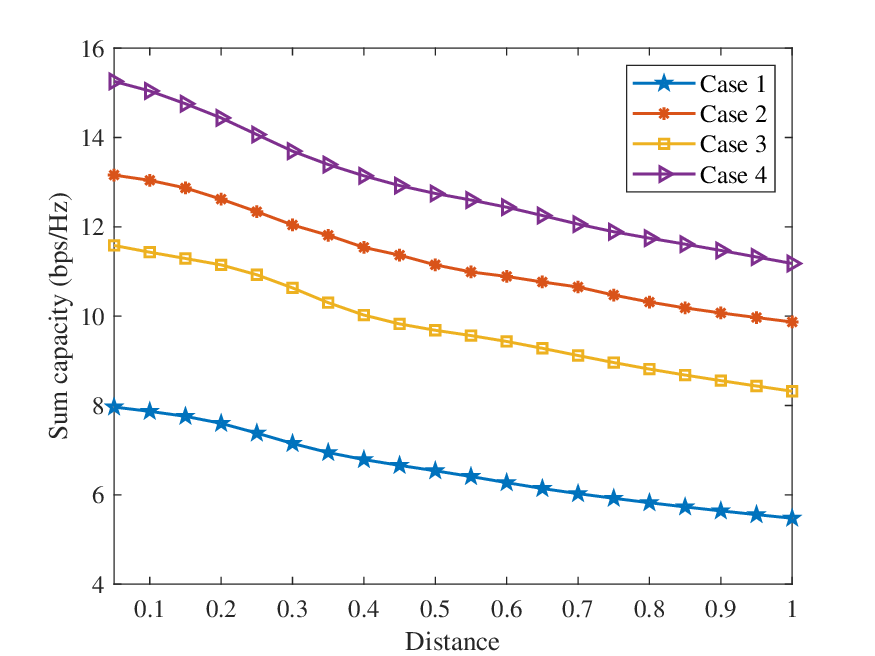}
\caption{The performance comparison of OAM communications with different OAM modes}
\label{fig:side:a}
\intextsep=1pt plus 3pt minus 1pt
\end{figure}
The capacity of different OAM mode combination is presented in Fig. \ref{fig:side:a}. It can be observed that the sum capacity is a monotonic decreasing function of the transmission distance. In addition, the proposed OAM communications with case 1 obtains the highest sum capacity. This is due to the fact that the array antenna combination of case 1 has the largest number of independent OAM channels. In other words, vortex wave beams with different OAM modes are equivalent to multiple independent OAM channels, which provides high sum capacity.

\begin{figure}[thp]
\centering
\includegraphics[height=2.8in,width=9cm]{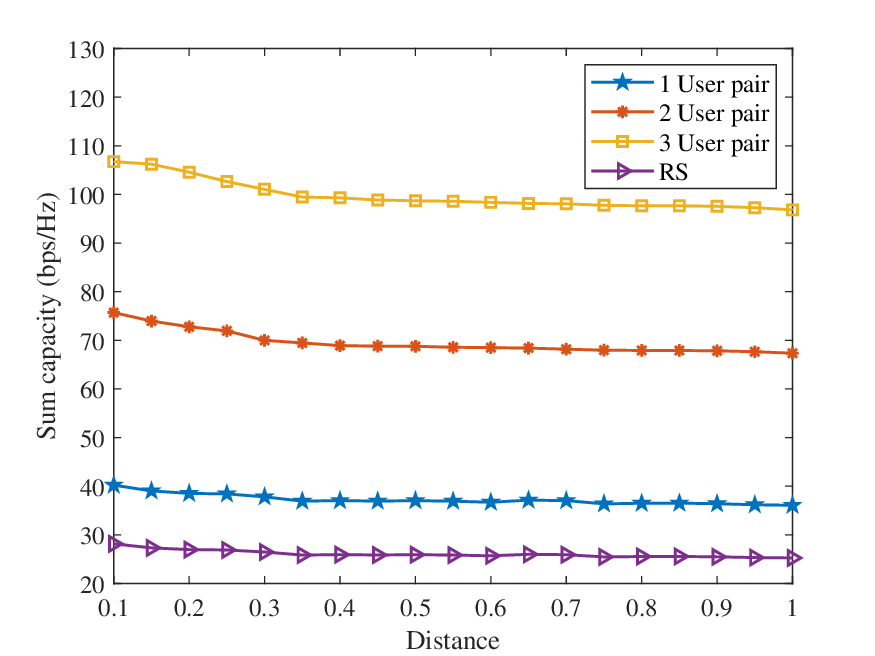}
\caption{The capacity performance of OAM communications with multiple OAM modes}
\label{fig:side:b}
\intextsep=1pt plus 3pt minus 1pt
\end{figure}
Figure \ref{fig:side:b} plots the sum capacity versus the transmission distance with multiple OAM modes.
%Moreover, the impact of different values of normalized $d$ (normalized distance between BS and user) on the user capacities also analyzed in Fig. 4.
We can see that the sum capacity monotonically decreases with the increase of transmission distance.
Furthermore, compared to the traditional communications with RS, the proposed OAM communications with different user pairs can obtain higher capacity. This is because the proposed OAM communications with multiple OAM modes can simultaneously transmit multiple signals by using multiple independent OAM channels. In addition, the channel condition between BS and ${\rm{U}}{{\rm{D}}}$ decreases slowly as the transmission distance increases.

\begin{figure}[thp]
\centering
\includegraphics[height=2.8in,width=9cm]{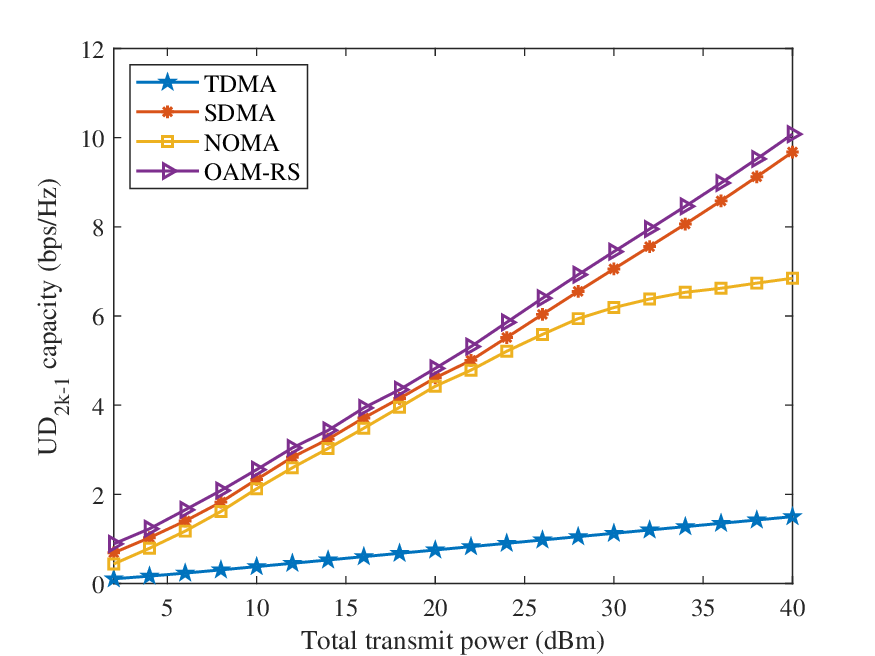}
\caption{The variation of ${\rm{U}}{{\rm{D}}_{2k-1}}$ capacity with total transmit power}
\label{fig:side:c}
\intextsep=1pt plus 3pt minus 1pt
\end{figure}

\begin{figure}[thp]
\centering
\includegraphics[height=2.8in,width=9cm]{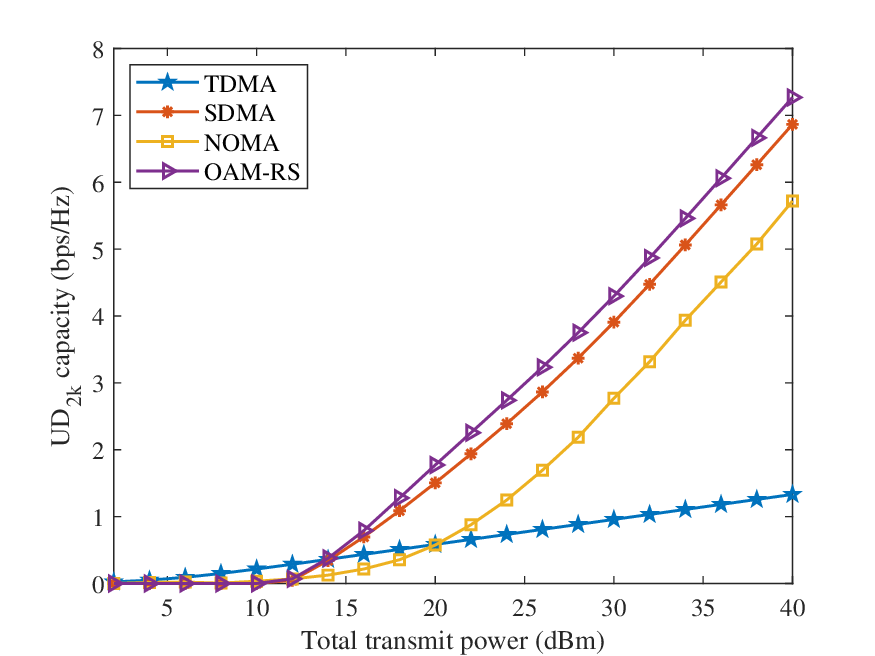}
\caption{The ${\rm{U}}{{\rm{D}}_{2k}}$ capacity of different wireless communications versus total transmit power}
\label{fig:side:d}
\intextsep=1pt plus 3pt minus 1pt
\end{figure}
The variation of ${\rm{U}}{{\rm{D}}_{2k-1}}$ capacity with the transmission distance is illustrated in Fig. \ref{fig:side:c}.
We can see that the sum capacity is a monotonic increasing function of the total transmit power. Furthermore, compared with the traditional wireless communications, OAM-MIMO communications with RS can achieve the
highest capacity. This is due to the utilization of multiple OAM modes,
which enables different vortex wave beams to transmit multiple signals independently.
Figure \ref{fig:side:d} demonstrates the ${\rm{U}}{{\rm{D}}_{2k}}$ capacities of different wireless communications. From Fig. \ref{fig:side:d}, it can be observed that TDMA communications can achieve higher capacity than the other three wireless communications if the total transmit power is relatively small. When the total transmit power is large, OAM-MIMO communications with RS can obtain the highest capacity among the four wireless communications.

%\begin{figure}[thp]
%\centering
%\includegraphics[height=2.6in,width=9cm]{sum.eps}
%\caption{Sum capacity regarding transmit power.}
%\label{fig:side:c}
%\intextsep=1pt plus 3pt minus 1pt
%\end{figure}
%Figure 3 demonstrates the variation of sum capacity for the four wireless communications. the sum capacity of RBVW communications increases with the increase of the transmit power. Moreover, the sum capacity for the RBVW communications has higher capacity than the other communications.

\begin{figure}[thp]
\centering
\includegraphics[height=2.8in,width=9cm]{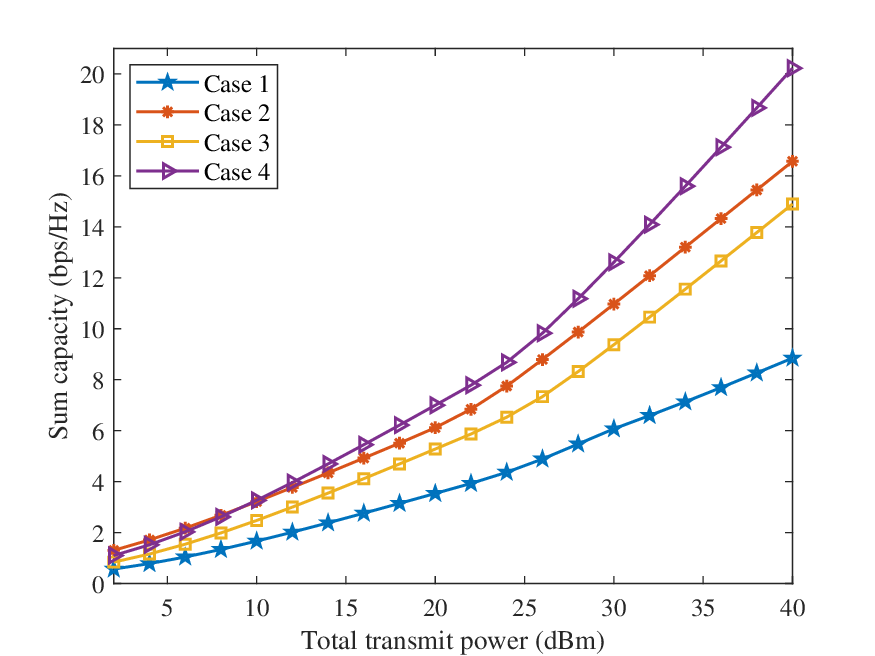}
\caption{The capacities of different cases under various total transmit power}
\label{fig:side:f}
\intextsep=1pt plus 3pt minus 1pt
\end{figure}
Figure \ref{fig:side:f} illustrates the performance comparison of OAM communications with multiple OAM modes. We can observe that the proposed OAM communications with case 1 achieves higher sum capacity than the other three cases. This is because the array antenna combination of case 1 produces the largest number of independent channels. Multiple vortex wave beams with various OAM modes simultaneously transfer multiple OAM signals to the user pairs by utilizing multiple independent OAM channels. Each OAM channel carries different OAM signal to the user pair. The more the OAM modes are, the higher the sum capacity is.

\section{Conclusions}

To enhance the privacy performance, we integrated RS technique into OAM communications and constructed the transmission framework of the downlink OAM-MIMO communications with RS. Different user pairs employed different OAM modes and users in the same user pair used RS technique to obtain the information. Based on concentric UCA, the OAM-MIMO channel model and the capacity were derived for the downlink OAM-MIMO communications with RS. The simulation experiments were conducted to show that the proposed precoding optimization algorithm can achieve higher sum capacity compared with the traditional schemes.

\begin{IEEEbiography}[{\includegraphics[width=0.9in,height=1.25in]{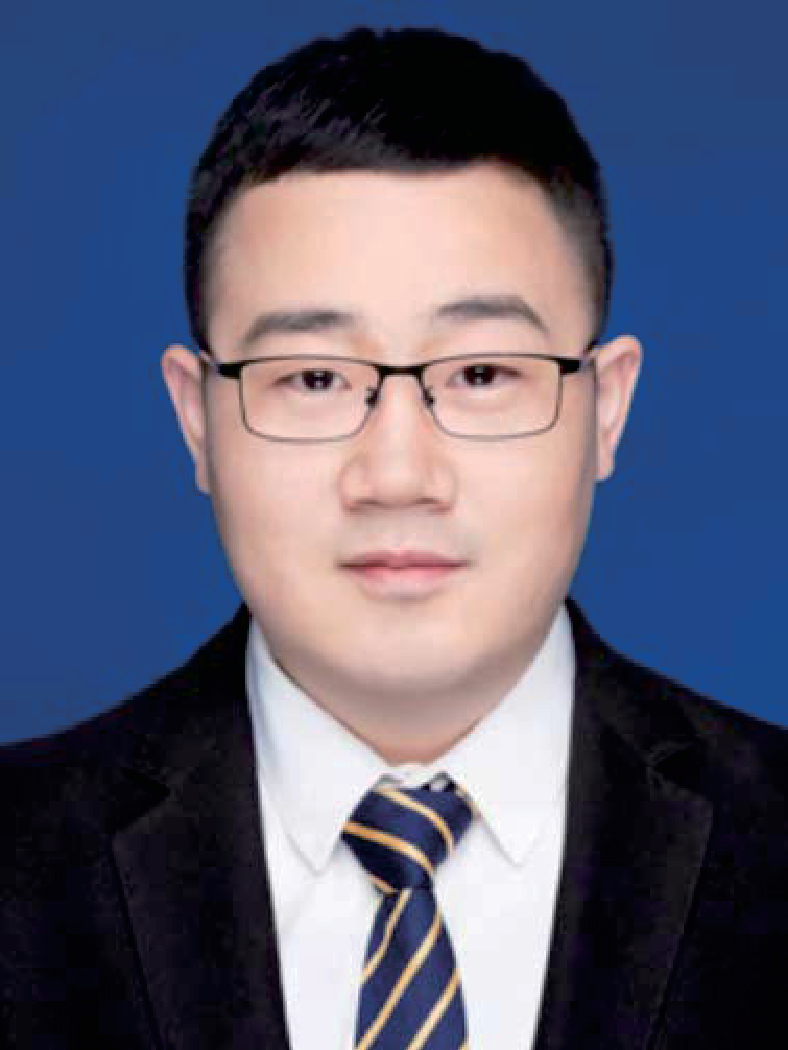}}]{Ruirui Chen}
received the Ph.D. degree in military communication from Xidian University, Xi'an, China, in 2018. He is currently a lecturer with the School of Information and Control Engineering, China University of Mining and Technology, Xuzhou, China. He is also a master supervisor of electronic information in China University of Mining and Technology. His research interests focus on vortex electromagnetic wave communications, MIMO communications and UAV communications. Email: rrchen@cumt.edu.cn.
\end{IEEEbiography}

\begin{IEEEbiography}[{\includegraphics[width=0.9in,height=1.25in]{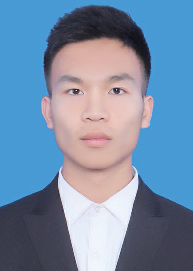}}]{Jinyang Lin}
received the B.S. degree in information and communication engineering from China University of Mining and Technology, Xuzhou, China, in 2021. He is currently working towards the M.S. degree with the School of Information and Control Engineering, China University of Mining and Technology, Xuzhou, China. His current research interests focus on in vortex wave communication. Email: linjinyang@cumt.edu.cn.
\end{IEEEbiography}
\vspace{-100 mm}
\begin{IEEEbiography}[{\includegraphics[width=0.9in,height=1.25in]{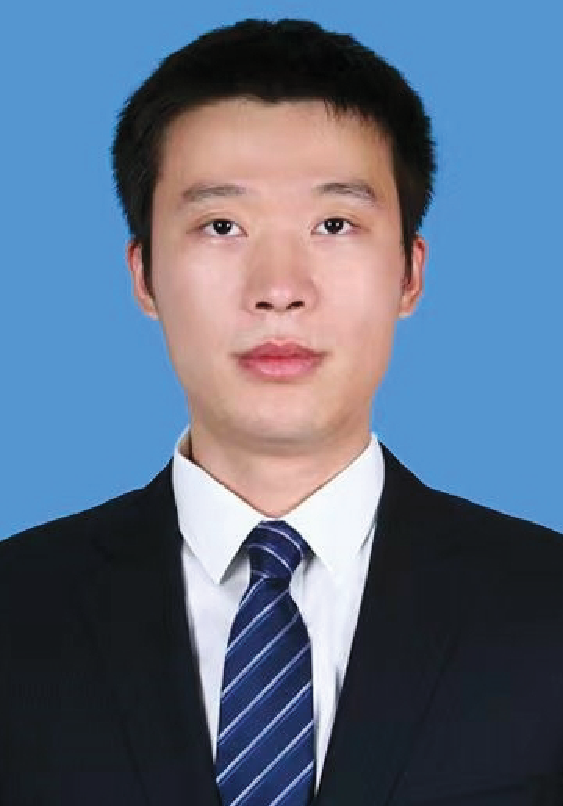}}]{Beibei Zhang}
received the M.S. degree in information and communication engineering from China University of Mining and Technology, Xuzhou, China, in 2011. He is currently a Senior Engineer with Jiangsu Automation Research Institute, Lianyungang, China and also working towards the Ph.D. degree with the School of Information and Control Engineering, China University of Mining and Technology, Xuzhou, China. His research interests include wireless sensor networks and vortex wave communications. Email: lb15060029@cumt.edu.cn.
\end{IEEEbiography}
\vspace{-100 mm}

\begin{IEEEbiography}[{\includegraphics[width=0.9in,height=1.25in]{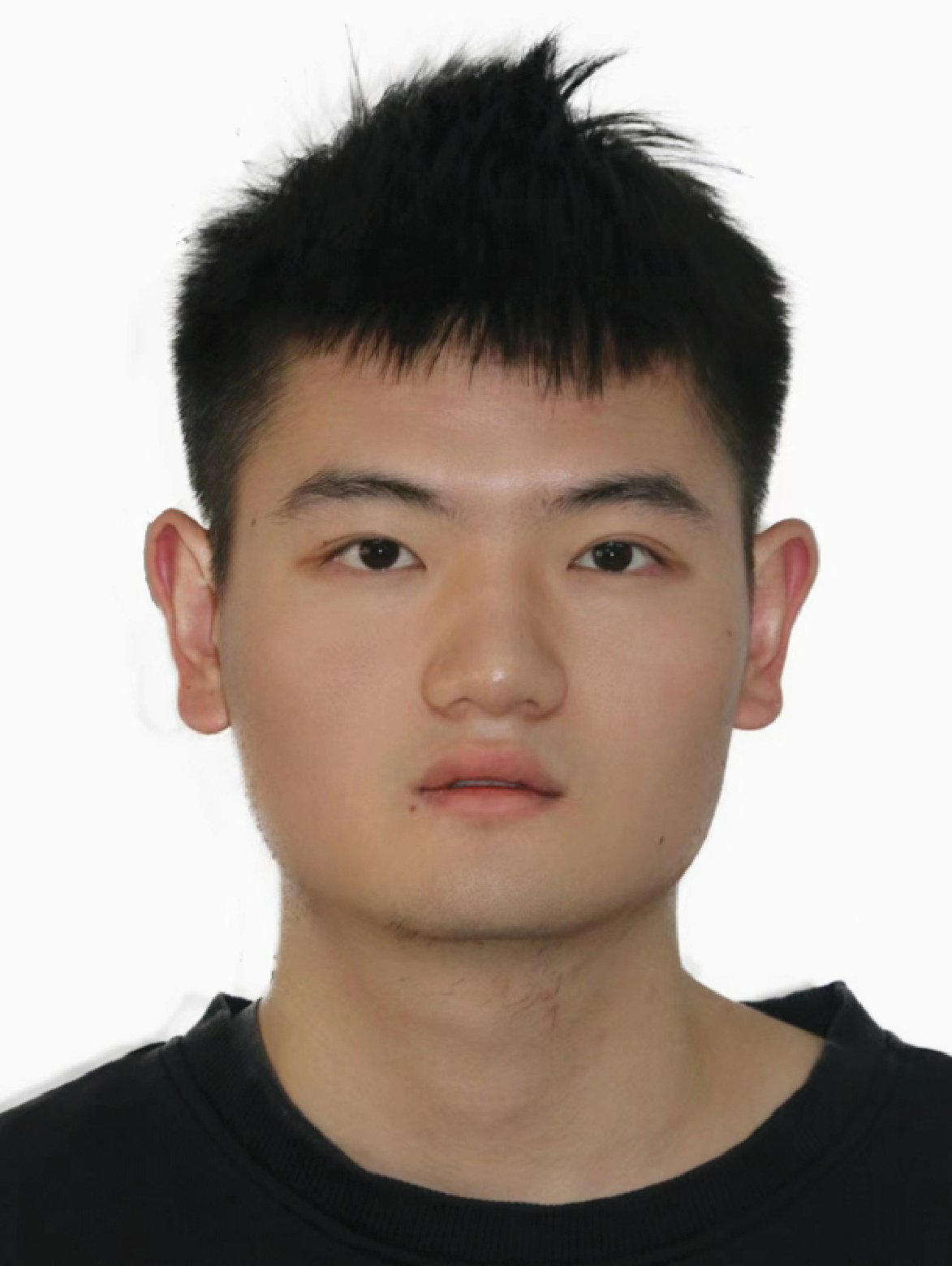}}]{Yu Ding}
received the B.S. degree in information and communication engineering from China University of Mining and Technology, Xuzhou, China, in 2022. He is currently working towards the M.S. degree with the School of Information and Control Engineering, China University of Mining and Technology, Xuzhou, China. His current research interests focus on in UAV communications and vortex wave communication. Email: yding@cumt.edu.cn.
\end{IEEEbiography}
\vspace{-100 mm}
\begin{IEEEbiography}[{\includegraphics[width=0.9in,height=1.25in]{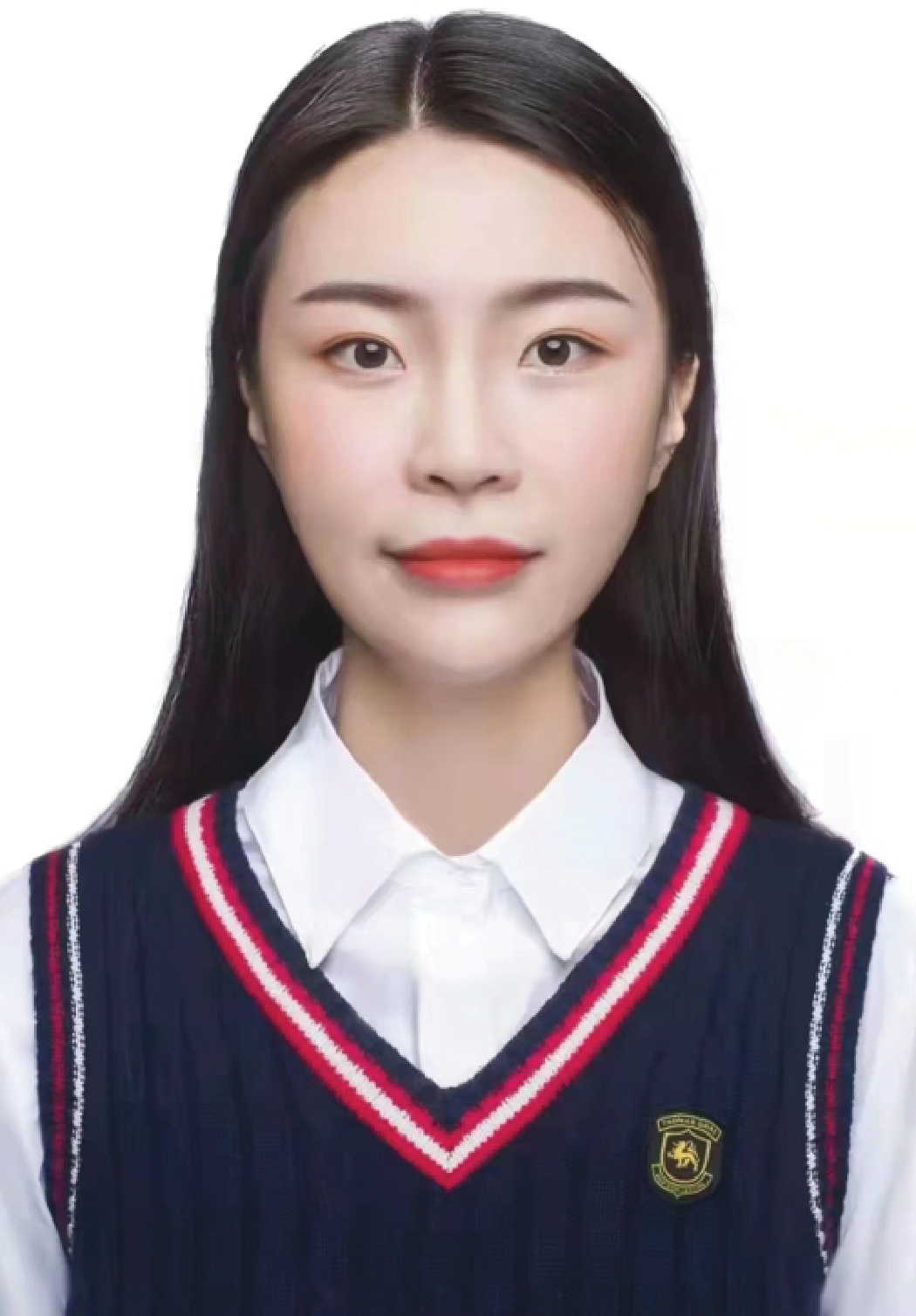}}]{Keyue Xu}
received the B.S. degree in communication engineering from Suzhou University of Science and Technology, Suzhou, China, in 2022. She is currently working towards the M.S. degree with the School of Information and Control Engineering, China University of Mining and Technology, Xuzhou, China. Her current research interests focus on in intelligent communications and vortex wave communication. Email: kyxu@cumt.edu.cn.
\end{IEEEbiography}


\begin{thebibliography}{99}% more than 9 --> 99 / less than 10 --> 9
\bibitem{1}
W. Cheng, W. Zhang, H. Jing, et al, ``Orbital angular momentum for wireless communications," \emph{IEEE Wireless Communications}, vol. 26, no. 1, pp. 100-107, Feb. 2019.

\bibitem{2a}
H. Jing, W. Cheng, W. Zhang, et al, ``Optimal UCA design for OAM based wireless backhaul transmission," \emph{ICC 2020 - 2020 IEEE International Conference on Communications (ICC)}, pp. 1-6, 2020.

\bibitem{2}
A. Trichili, K. Park, M. Zghal, et al, ``Communicating using spatial mode multiplexing: potentials, challenges, and perspectives," \emph{IEEE Communications Surveys \& Tutorials}, vol. 21, no. 4, pp. 3175-3203, 2019.

\bibitem{3a}
R. Lyu, W. Cheng, B. Shen, et al, ``OAM-SWIPT for IoE-driven 6G," \emph{IEEE Communications Magazine}, vol. 60, no. 3, pp. 19-25, Mar. 2022.

\bibitem{3}
Y. Lei, Y. Yang, Y. Wang, et al, ``Throughput performance of wireless multiple-input multiple-output systems using OAM antennas," \emph{IEEE Wireless Communications Letters}, vol. 10, no. 2, pp. 261-265, Feb. 2021.

\bibitem{4}
R. Chen, H. Zhou, M. Moretti, et al., ``Orbital angular momentum waves: generation, detection, and emerging applications," \emph{IEEE Communications Surveys \& Tutorials}, vol. 22, no. 2, pp. 840-868, 2020.

\bibitem{5}
R. Lyu, W. Cheng and W. Zhang, ``Modeling and performance analysis of OAM-NFC systems," \emph{IEEE Transactions on Communications}, vol. 69, no. 12, pp. 7986-8001, Dec. 2021.

\bibitem{6}
H. Jing, W. Cheng, Z. Li, et al., ``Concentric UCAs based low-order OAM for high capacity in radio vortex wireless communications," \emph{Journal of Communications and Information Networks}, vol. 3, no. 4, pp. 85-100, Dec. 2018.
%%%%%%%%%%%%%%%%%%%%%%%%%%%%%%%%%%%%%%%%%%%%%%%%%%%%%%%%%%%%%%%%%%%%%%%%%%%%%%%%%%%%%%%%%%%%%%%%%%%%%%%%%%
\bibitem{7}
W. Long, R. Chen, M. Moretti, et al., ``Joint spatial division and coaxial multiplexing for downlink multi-user OAM wireless backhaul," \emph{IEEE Transactions on Broadcasting}, vol. 67, no. 4, pp. 879-893, Dec. 2021.

\bibitem{8}
A. Amin and S. Shin, ``Channel capacity analysis of non-orthogonal multiple access with OAM-MIMO system," \emph{IEEE Wireless Communications Letters}, vol. 9, no. 9, pp. 1481-1485, Sep. 2020.

\bibitem{9}
H. Yin and H. Liu, ``Performance of space-division multiple-access (SDMA) with scheduling," \emph{IEEE Transactions on Wireless Communications}, vol. 1, no. 4, pp. 611-618, Oct. 2002.

\bibitem{10}
S. Thoen, L. Perre, M. Engels, et al., ``Adaptive loading for OFDM/SDMA-based wireless networks," \emph{IEEE Transactions on Communications}, vol. 50, no. 11, pp. 1798-1810, Nov. 2002.

\bibitem{11}
L. Dai, B. Wang, Y. Yuan, et al., ``Non-orthogonal multiple access for 5G: solutions, challenges, opportunities, and future research trends," \emph{IEEE Communications Magazine}, vol. 53, no. 9, pp. 74-81, Sep. 2015.

\bibitem{12}
N. Zhao, D. Li, M. Liu, et al., ``Secure transmission via joint precoding optimization for downlink MISO NOMA," \emph{IEEE Transactions on Vehicular Technology}, vol. 68, no. 8, pp. 7603-7615, Aug. 2019.

\bibitem{13}
R. Chen, F. Cheng, J. Lin, et al., ``Performance analysis of rate splitting multiple access based vortex wave communications," \emph{IEEE Wireless Communications Letters}, vol. 11, no. 8, pp. 1570-1574, Aug. 2022.

\bibitem{14}
D. Yu, J. Kim and S. Park,  ``An efficient rate-splitting multiple access scheme for the downlink of C-RAN systems," \emph{IEEE Wireless Communications Letters}, vol. 8, no. 6, pp. 1555-1558, Dec. 2019.

\bibitem{15}
Y. Mao, B. Clerckx and V. Li, ``Energy efficiency of rate-splitting multiple access, and performance benefits over SDMA and NOMA," \emph{2018 15th International Symposium on Wireless Communication Systems (ISWCS)}, pp. 1-5, 2018.

\bibitem{16}
Z. Lin, M. Lin, T. Cola, et al., ``Supporting IoT with rate-splitting multiple access in satellite and aerial-integrated networks," \emph{IEEE Internet of Things Journal}, vol. 8, no. 14, pp. 11123-11134, Jul. 2021.

\bibitem{17}
J. Zhang, B. Clerckx, J. Ge, et al., ``Cooperative rate splitting for MISO broadcast channel with user relaying, and performance benefits over cooperative NOMA," \emph{IEEE Signal Processing Letters}, vol. 26, no. 11, pp. 1678-1682, Nov. 2019.

\bibitem{18}
D. Li, Z. Yang, N. Zhao, et al., ``Precoding optimization assisted secure transmission for rate-splitting multiple access," \emph{ICC 2022 - IEEE International Conference on Communications}, pp. 673-678, Aug. 2022.

\bibitem{19}
Y. Mao, B. Clerckx and V. Li, ``Rate-splitting for multi-antenna non-orthogonal unicast and multicast transmission: Spectral and energy efficiency analysis," \emph{IEEE Transactions on Communications}, vol. 67, no. 12, pp. 8754-8770, Dec. 2019.



%%%%%%%%%%%%%%%%%%%%%%%%%%%%%%%%%%%%%%%%%%%%%%%%%%%%%%%%%%%%%%%%%%%%%%%%%%%%%%%%%%%%%%%%%%%%%%%%%%%%%%%%%%

\bibitem{20}
Te Han and K. Kobayashi, ``A new achievable rate region for the interference channel," \emph{IEEE Transactions on Information Theory}, vol. 27, no. 1, pp. 49-60, Jan. 1981.

\bibitem{21}
Y. Mao, B. Clerckx, and V. Li, ``Rate-splitting multiple access for downlink communication systems: bridging, generalizing, and outperforming SDMA and NOMA," \emph{EURASIP journal on wireless communications and networking}, vol. 2018, no. 1, p. 133, May 2018.

\bibitem{22}
B. Clerckx, Y. Mao, R. Schober, et al., ``Rate-splitting unifying SDMA, OMA, NOMA, and multicasting in MISO broadcast channel: a simple two-user rate analysis," \emph{IEEE Wireless Communications Letters}, vol. 9, no. 3, pp. 349-353, Mar. 2020.

\bibitem{23}
B. Clerckx, H. Joudeh, C. Hao, et al., ``Rate splitting for MIMO wireless networks: a promising PHY-layer strategy for LTE evolution,"  \emph{IEEE Communications Magzine}, vol. 54, no. 5, pp. 98-105, May 2016.

\bibitem{24}
H. Joudeh and B. Clerckx, ``Rate-splitting for max-min fair multigroup multicast beamforming in overloaded systems," \emph{IEEE Transactions on Wireless Communications}, vol. 16, no. 11, pp. 7276-7289, Nov. 2017.

\bibitem{25}
H. Chen, D. Mi, B. Clerckx, et al., ``Joint power and subcarrier allocation optimization for multigroup multicast systems with rate splitting," \emph{IEEE Transactions on Vehicular Technology}, vol. 69, no. 2, pp. 2306-2310, Feb. 2020.

\bibitem{26}
M. Dai, B. Clerckx, D. Gesbert, et al., ``A rate splitting strategy for massive MIMO with imperfect CSIT," \emph{IEEE Transactions on Wireless Communications}, vol. 15, no. 7, pp. 4611-4624, Jul. 2016.

\bibitem{27}
O. Tervo, L. Trant, S. Chatzinotas, et al., ``Multigroup multicast beamforming and antenna selection with ratesplitting in multicell systems," \emph{2018 IEEE 19th International Workshop on Signal Processing Advances in Wireless Communications (SPAWC)}, pp. 1-5, Jun. 2018.

\bibitem{28}
S. Christensen, R. Agarwal, E. De Carvalho, et al., ``Weighted sum-rate maximization using weighted MMSE for MIMOBC beamforming design," \emph{IEEE Transactions on Wireless Communications}, vol. 7, no. 12, pp. 4792-4799, Dec. 2008.

\bibitem{29}
M. Vazquez, M. Caus, and A. Perez-Neira, ``Rate splitting for mimo multibeam satellite systems," \emph{WSA 2018: 22nd International ITG Workshop on Smart Antennas}, pp. 1-6, Mar. 2018.

\bibitem{30}
H. Joudeh and B. Clerckx, ``Robust transmission in downlink multiuser MISO systems: a rate-splitting approach," \emph{IEEE Transactions on Signal Processing}, vol. 64, no. 23, pp. 6227-6242, Dec. 2016.

%%%%%%%%%%%%%%%%%%%%%%%%%%%%%%%%%%%%%%%%%%%%%%%%%%%%%%%%%%%%%%%%%%%%%%%%%%%%%%%%%%%%%%%%%%%%%%%%%%%%%%%%%%

\bibitem{31}
Z. Zhang, Y. Xiao, Z. Ma, et al, ``6G wireless networks: vision, requirements, architecture, and key technologies," \emph{IEEE Vehicular Technology Magazine}, vol. 14, no. 3, pp. 28-41, Sept. 2019.

\bibitem{32}
P. Yang, Y. Xiao, M. Xiao, et al, ``6G wireless communications: vision and potential techniques," \emph{IEEE Network}, vol. 33, no. 4, pp. 70-75, Aug. 2019.

\bibitem{33}
G. Xie, L. Li, Y. Ren, et al, ``Performance metrics and design considerations for a free-space optical orbital-angular-momentum-multiplexed communication link," \emph{Optica}, vol. 2, no. 4, pp. 357-365, 2015.

\bibitem{34}
Y. Zhang, W. Feng and N. Ge, ``On the restriction of utilizing orbital angular momentum in radio communications," \emph{2013 8th International Conference on Communications and Networking in China (CHINACOM)}, pp. 271-275 2013.

\bibitem{35}
Z. Tian, R. Chen, W. Long, et al, ``Broadband beam steering for misaligned multi-mode OAM communication systems," \emph{Journal of Systems Engineering and Electronics}, vol. 32, no. 4, pp. 779-788, Aug. 2021.

\bibitem{36}
W. Yu, B. Zhou, Z. Bu, et al, ``UCA based OAM beam steering with high mode isolation," \emph{IEEE Wireless Communications Letters}, vol. 11, no. 5, pp. 977-981, May 2022.

\bibitem{37}
R. Chen, H. Xu, M. Moretti, et al, ``Beam steering for the misalignment in UCA-based OAM communication systems," \emph{IEEE Wireless Communications Letters}, vol. 7, no. 4, pp. 582-585, Aug. 2018.

\bibitem{38}
R. Chen, W. Long, X. Wang, et al, ``Multi-mode OAM radio waves: generation, angle of arrival estimation and reception with UCAs," \emph{IEEE Transactions on Wireless Communications}, vol. 19, no. 10, pp. 6932-6947, Oct. 2020.

\bibitem{39}
Y. Xiu, Y. Wang, P. Shi, et al, ``Maximization of millimeter-wave LoS OAM link using beam steering and partial receiving," \emph{2022 16th European Conference on Antennas and Propagation (EuCAP)},  pp. 1-5, 2022.

\bibitem{40}
M. Klemes, H. Boutayeb and F. Hyjazie, ``Orbital angular momentum (OAM) modes for 2-D beam-steering of circular arrays," \emph{2016 IEEE Canadian Conference on Electrical and Computer Engineering (CCECE)}, pp. 1-5, 2016.

\bibitem{41}
S. Zheng, Y. Chen, Z. Zhang, et al, ``Realization of beam steering based on plane spiral orbital angular momentum wave," \emph{IEEE Transactions on Antennas and Propagation}, vol. 66, no. 3, pp. 1352-1358, Mar. 2018.

\bibitem{42}
R. Chen, M. Zou, X. Wang, et al, ``Generation and beam steering of arbitrary-order OAM with time-modulated circular arrays," \emph{IEEE Systems Journal}, vol. 15, no. 4, pp. 5313-5320, Dec. 2021.

\bibitem{43}
S. Yu, N. Kou, J. Jiang, et al, ``Beam steering of orbital angular momentum vortex waves with spherical conformal array," \emph{IEEE Antennas and Wireless Propagation Letters}, vol. 20, no. 7, pp. 1244-1248, Jul. 2021.

\bibitem{44}
Q. Song, Y. Wang, K. Liu, et al. ``Beam steering for OAM beams using time modulated circular arrays," \emph{Electronics Letters}, vol. 54, no. 17, pp. 1017-1018, 2018.

\bibitem{45}
N. Qasem, A. Alamayreh, J. Rahhal, ``Beam steering using OAM waves generated by a concentric circular loop antenna array," \emph{Wireless Networks}, vol. 27, no. 4, pp. 2431-2440, 2021.

\bibitem{46}
K. Shen and W. Yu, ``Fractional programming for communication systems-part I: power control and beamforming," \emph{ IEEE Transactions on Signal Processing}, vol. 66, no. 10, pp. 2616-2630, May, 2018.
\end{thebibliography}
\end{document}